\PassOptionsToPackage{table,dvipsnames}{xcolor}
\documentclass[sigconf]{acmart}
\copyrightyear{2025}
\acmYear{2025}
\setcopyright{acmlicensed}\acmConference[HPDC '25]{The 34th International
Symposium on High-Performance Parallel and Distributed Computing}{July
20--23, 2025}{Notre Dame, IN, USA}
\acmBooktitle{The 34th International Symposium on High-Performance Parallel
and Distributed Computing (HPDC '25), July 20--23, 2025, Notre Dame, IN,
USA}
\acmDOI{10.1145/3731545.3731577}
\acmISBN{979-8-4007-1869-4/2025/07}

\usepackage{graphicx}
\usepackage{subcaption}
\usepackage{multirow}

\begin{document}

%%
%% The "title" command has an optional parameter,
%% allowing the author to define a "short title" to be used in page headers.
\title[TSUE]{TSUE: A Two-Stage Data Update Method for an Erasure Coded Cluster File System}

%%
%% The "author" command and its associated commands are used to define
%% the authors and their affiliations.
%% Of note is the shared affiliation of the first two authors, and the
%% "authornote" and "authornotemark" commands
%% used to denote shared contribution to the research.

\settopmatter{authorsperrow=4}

\author{Zheng Wei}
\affiliation{%
\small
 \institution{Institute of Computing Technology, Chinese Academy of Sciences}
 \city{Beijing}
 \country{China}}

\author{Jing Xing}
\authornote{Corresponding author: Jing Xing (xingjing@ncic.ac.cn).}
\affiliation{%
\small
 \institution{Institute of Computing Technology, Chinese Academy of Sciences}
 \city{Beijing}
 \country{China}}

\author{Yida Gu}
\affiliation{%
\small
 \institution{Institute of Computing Technology, Chinese Academy of Sciences}
 \city{Beijing}
 \country{China}}

\author{Wenjing Huang}
\affiliation{%
\small
 \institution{Institute of Computing Technology, Chinese Academy of Sciences}
 \city{Beijing}
 \country{China}}

\author{Dong Dai}
\affiliation{%
\small
 \institution{University of Delaware}
 \city{Newark}
 \state{DE}
 \country{USA}
 }

\author{Guangming Tan}
\affiliation{%
\small
 \institution{Institute of Computing Technology, Chinese Academy of Sciences}
 \city{Beijing}
 \country{China}}

\author{Dingwen Tao}
\affiliation{%
 \small
 \institution{Institute of Computing Technology, Chinese Academy of Sciences}
 \city{Beijing}
 \country{China}}
%%
%% By default, the full list of authors will be used in the page
%% headers. Often, this list is too long, and will overlap
%% other information printed in the page headers. This command allows
%% the author to define a more concise list
%% of authors' names for this purpose.
\renewcommand{\shortauthors}{Wei et al.}

%\settopmatter{printacmref=false}
%\setcopyright{none}
%\renewcommand\footnotetextcopyrightpermission[1]{}

%\pagestyle{plain}

%%
%% The abstract is a short summary of the work to be presented in the
%% article.
\begin{abstract}
  Compared to replication-based storage systems, erasure-coded storage incurs significantly higher overhead during data updates. To address this issue, various parity logging methods have been proposed. Nevertheless, due to the long update path and substantial amount of random I/O involved in erasure code update processes, the resulting long latency and low throughput often fail to meet the requirements of high performance applications. To this end, we propose a two-stage data update method called TSUE. TSUE divides the update process into a synchronous stage that records updates in a data log, and an asynchronous stage that recycles the log in real-time. TSUE effectively reduces update latency by transforming random I/O into sequential I/O, and it significantly reduces recycle overhead by utilizing a three-layer log and the spatio-temporal locality of access patterns. In SSDs cluster, TSUE significantly improves update performance, achieving improvements of 7.6$\times$ under Ali-Cloud trace, 5$\times$ under Ten-Cloud trace, while it also extends the SSD's lifespan by up to 13$\times$ through reducing the frequencies of reads/writes and of erase operations.
\end{abstract}

%%
%% The code below is generated by the tool at http://dl.acm.org/ccs.cfm.
%% Please copy and paste the code instead of the example below.
%%
    


%%
%% Keywords. The author(s) should pick words that accurately describe
%% the work being presented. Separate the keywords with commas.
%Spatio-temporal Locality, DataLog, DeltaLog, ParityLog, 3-Layer log structure, Log Pool Structure based on FIFO, Memory, Real-Time Recycle
%% A "teaser" image appears between the author and affiliation
%% information and the body of the document, and typically spans the
%% page.

%%
%% This command processes the author and affiliation and title
%% information and builds the first part of the formatTSUEted document.

\maketitle  

\section{Introduction}
\begin{comment}
   ACM's consolidated article template, introduced in 2017, provides a
consistent \LaTeX\ style for use across ACM publications, and
incorporates accessibility and metadata-extraction functionality
necessary for future Digital Library endeavors. Numerous ACM and
SIG-specific \LaTeX\ templates have been examined, and their unique
features incorporated into this single new template.

If you are new to publishing with ACM, this document is a valuable
guide to the process of preparing your work for publication. If you
have published with ACM before, this document provides insight and
instruction into more recent changes to the article template.

The ``\verb|acmart|'' document class can be used to prepare articles
for any ACM publication --- conference or journal, and for any stage
of publication, from review to final ``camera-ready'' copy, to the
author's own version, with {\itshape very} few changes to the source. 
\end{comment}
    
\label{Intro}
\begin{comment}
Erasure coding is widely acknowledged as a highly efficient technique in distributed storage systems due to its ability to minimize storage overhead. In comparison to the replication mechanism, erasure coding achieves lower storage redundancy while maintaining an equivalent level of reliability. For example, the replica mechanism is usually used to ensure reliability, with three replicas adopted, resulting in a 200\% storage overhead. However, RS(4,2)\footnote{RS(K,M): In the erasure code file system, files are segmented into continuous fixed data blocks of 64 MB, 128 MB, or 256 MB. The Reed-Solomon algorithm generates M parity blocks from K data blocks. These K data blocks and M parity blocks constitute a stripe with a total size of $K+M$. When no more than M blocks within a stripe are lost due to devices failure, the lost data can be reconstructed using the invertible full-rank matrix property of the Reed-Solomon algorithm.} encoding can ensure the same reliability with only a 50\% storage overhead. In fact, there are extreme environments that use RS(128,4) \cite{ref28} encoding to maintain data reliability, with storage overhead as low as 3\%. In recent years, erasure coding has emerged as a mature and extensively adopted alternative. It has been embraced by most open-source systems and commercial storage systems including GoogleFS \cite{ref1}, Microsoft Windows Azure Storage \cite{ref11}, Amazon \cite{ref15}, Facebook \cite{ref16}, GlusterFS \cite{ref20},  CephFS \cite{ref21}, Apache HDFS \cite{ref24} and Memcached \cite{ref28}.
\end{comment}

\textit{Motivation.} Erasure coding is widely acknowledged as a highly efficient technique in distributed storage systems due to its ability to minimize storage overhead. In comparison to the replication mechanism, erasure coding achieves lower storage redundancy while maintaining an equivalent level of reliability  \cite{ref4, ref10, ref12, ref13, ref14, ref18, ref19, ref23,ref25}. In recent years, erasure coding has emerged as a mature and extensively adopted alternative. It has been embraced by most open-source systems and commercial storage systems including GoogleFS \cite{ref1}, Microsoft Windows Azure Storage \cite{ref11}, Amazon \cite{ref15}, Facebook \cite{ref16}, GlusterFS \cite{ref20},  CephFS \cite{ref21}, Apache HDFS \cite{ref24} and Memcached \cite{ref28}.

The random-access overhead and the long update path in the update process of erasure coded file systems increases the update latency and decreases the update throughput. During the process of updating erasure coding, it is imperative to concurrently update both the data blocks and their corresponding parity blocks. Presently, a majority of commercial and open-source systems employ reconstruction update and incremental update mechanisms for data modifications. The reconstruction update operation requires reading the unchanged portion of the stripe over the network to form a complete data segment, thus enabling the recalculation of the parity blocks. This process incurs significant read and network overhead, as seen in systems like GlusterFS \cite{ref20} and CephFS \cite{ref21}. In contrast, incremental update only affects the updated data block and the relevant parity blocks, significantly reducing the read and write penalties associated with the update operation. 

However, the current incremental update mechanism faces several challenges. Therefore, it is necessary to design an erasure code update mechanism that ensures low latency, high throughput, high consistency, and high devices lifespan.

\textit{Limitation of state-of-art approaches.}
Current incremental update methods for erasure codes continue to suffer from some issues:

\textbf{High Update Latency:}
The update latency remains high due to the lengthy data update path and the inherent nature of random access operations during the update process. The potential for optimizing update latency is constrained by the in-place update of data blocks used in the incremental update mechanism, which necessitates a time-consuming read-write process to compute the data delta which is the difference between new data and original data as shown in Equation (\ref{eq2}). Furthermore, the involved access patterns are small-grained and random. On SSD devices, the read and write latency for random access is several times higher than that for sequential operations.

\textbf{Consistency Issues caused by ParityLog:}
The hysteretic recycle of parity logs negatively impacts data consistency during data recovery. Most incremental update methods postpone the recycle process of parity logs until the storage space reaches a certain threshold or data loss occurs. Prolonged log recycle may lead to secondary data loss, while log loss can result in data integrity issues, both of which contribute to data consistency issues.

\textbf{Low Throughput:}
Throughout the update process of erasure codes, read and write operations on both data blocks and parity blocks involve fine-grained random access. This access pattern significantly constrains the enhancement of update throughput. To meet the stringent access requirements of burst data and high-performance applications, it is essential to achieve high throughput.

\textbf{Low Lifespan:}
An update operation on an erasure code necessitates multiple overwrite operations for both data blocks and their corresponding parity blocks. On SSDs, a high volume of fine-grained, random overwrite operations can substantially degrade write performance and reduce the lifespan of storage units.

\textit{Key insights and contributions.}
To tackle the challenges of high update latency, weak consistency, low throughput and low lifespan in erasure coding, we propose TSUE, a two-stage update mechanism designed to optimize update workloads with temporal-spatial characteristics, reduce network traffic, and minimize SSD wear. 

In summary, we make the following contributions in this work: 

\begin{itemize}
    \item TSUE divides the update process into two stages: log appending and log recycling. The former is a synchronous front-end operation, and the latter is an asynchronous back-end operation. In the front-end processing, update data is directly appended to a data log using a sequential append approach, significantly reducing the perceived update latency for users. In the back-end processing, the log is recycled and merged with the original data to minimize any impact on data reading and recovery reliability. 

    \item TSUE employs a three-layer log structure for log management and recycles logs in parallel using a multi-layer log pipeline to enhance log recycling efficiency. The DataLog, DeltaLog, and ParityLog leverage the spatio-temporal characteristics of data access to reduce the number of requests and increase request granularity, thereby significantly improving overall recycling efficiency. Additionally, the DeltaLog exploits the spatial characteristics of data access to merge multiple data deltas for the same address across multiple data blocks within the same stripe, minimizing random-access workload and network traffic.

    \item We design a log pool structure for TSUE to facilitate concurrent log appending and recycling. Within the log unit, we propose a two-level index to handle requests that exhibit spatial-temporal locality. This index reduces random access by merging repeated requests, improves access granularity by combining small-grained requests. Additionally, it serves as a cache to enable high-speed data access.

    \item We implement TSUE and SOTA update methods (such as FO, PL, PLR, PARIX and CoRD) in an erasure-coded cluster file system, comparing them in various tests including Ten-Cloud Trace and Ali-Cloud Trace. Evaluation demonstrates that TSUE offers superior update performance. It also significantly enhances SSD lifespan by reducing erase writes and improving overall update efficiency.
\end{itemize}

\textit{Experimental methodology and artifact availability.} To verify the superiority of our approach, we conducted experiments using 16 nodes equipped with SSDs in the Chameleon cloud environment, using a 25 Gb/s inter-node network. For a fair comparison, we implemented FO, PL, PLR, PARIX, CoRD, and TSUE within our self-developed erasure code file system. To comprehensively evaluate update performance, we utilized trace data from Ali-Cloud and Ten-Cloud to conduct throughput tests under various Reed-Solomon (RS) encoding schemes, as well as contribution point analysis to highlight the key contributions of this work. We also analyzed the I/O load, overwrite load, and network traffic data generated by various update mechanisms. We will provide all necessary artifacts to reproduce the experimental results presented in this paper.

\textit{Limitations of the proposed approach.}
TSUE optimizes update performance by leveraging spatio-temporal access patterns and using a multi-level log structure to adjust data granularity and volume. In the RS(k,M) algorithm, TSUE is particularly suitable for environments with large M values. Test results show that as the M value decreases from 4 to 2, the performance advantage of TSUE gradually diminishes. Additionally, TSUE is highly effective for NAND granule-based SSD devices, where there is a significant performance gap between random and sequential read/write operations, as validated on the Chameleon platform. While we couldn't test high-end SSDs due to their cost, our approach remains beneficial when update requests exhibit temporal and spatial locality.

\section{Background and Motivation}

\label{Basics of EC updates}
Erasure coding generates $M$ parity blocks from $K$ fixed data blocks through matrix (Vandermonde Matrix or Cauchy Matrix) multiplication over a Galois Field \cite{ref29, ref30, ref41, ref42, ref43, ref22, ref33}, as shown in Equation (\ref{eq1}). These $K+M$ blocks are then distributed across various nodes within the storage system. The recovery of lost data relies on the reversibility of full-rank matrices. The matrix multiplication operations over a Galois Field are performed using either a Vandermonde matrix or a Cauchy matrix to generate the corresponding parity blocks, as illustrated in Equation (\ref{eq1}). In case of data loss, where up to $N$ blocks are lost ($N \leq M$), the missing blocks can be recovered by multiplying any $K$ surviving blocks with the inverse matrix of their corresponding encoding matrix.

\subsection{Existing Update Approaches}
The erasure code mechanism with low redundancy faces performance issues during update operations, which make up a large portion of write requests, mostly small-grained.  Statistical analysis of MSR Cambridge Trace \cite{ref2,ref34,ref35,ref36,ref37} reveals that approximately 60\% of updates across all volumes are smaller than 4KB, and 90\% are smaller than 16KB, with over 90\% of write requests being updates \cite{ref2}. In the Ali-Cloud trace analysis \cite{ref48}, 75\% of the requests are identified as update requests. Among these, 60\% have a size not exceeding 16 KB, with 46\% specifically sized at 4 KB. In the statistical analysis of Ten-Cloud traces \cite{ref49}, 69\% of the requests are update requests. Among these, 88\% on average have a size not exceeding 16 KB, with 69\% on average specifically sized at 4 KB.

Erasure coding updates can be classified into two main types: reconstruct approaches and incremental approaches. 

\subsubsection{Reconstruct Approaches:}
Given the random nature and large volume of update data, rebuilding a complete stripe within limited memory space and time constraints becomes challenging. For instance, ECWide \cite{ref27} emphasizes that the maximum $k$ value for RS encoding within clusters is set to 128.  Each 4KB update requires at least 512KB of memory for caching in the stripe to compute parity data chunks. Thus, processing 1 million update requests simultaneously would require a minimum of 512GB of memory, leading to excessive memory usage. Besides, the reconstructive update method introduces significant network and random-access overhead, making it more suitable for scenarios involving sequential updates that cover the entire data portion of a stripe. 
 
{
\small
\begin{align}
    \left[\begin{matrix} P_1  \\ P_2 \\ P_3 \\ \vdots \\ P_M \end{matrix} \right]  = 
    \left[\begin{matrix}
            \partial_{11} & \partial_{12} & \dotsc & \partial_{1K} \\
            \partial_{21} & \partial_{22} & \dotsc & \partial_{2K} \\
            \partial_{31} & \partial_{32} & \dotsc & \partial_{3K} \\
            \vdots & \vdots & \ddots & \vdots \\
            \partial_{M1} & \partial_{M2} & \dotsc & \partial_{MK}
        \end{matrix} \right] \cdot 
        \left[\begin{matrix}  D_1  \\ D_2 \\ D_3 \\ \vdots \\  D_K \end{matrix} \right] 
        \label{eq1}
\end{align}
\begin{align}
    P_1^n &= P_1^{n-1}+ \partial_{11}\ast(D_1^n-D_1^{n-1})     \label{eq2} \\
    P_1^n &= P_1+ \partial_{11}\ast\left(\left(D_1^n-D_1^{n-1}\right)+\ldots+\left(D_1^1-\left.\ D_1\right)\right.\right) \label{eq3} \\
    P_1^n &= P_1+ \partial_{11}\ast\left(\left(D_1^n-\left.\ D_1\right)\right.\right)   \label{eq4} 
\end{align}
\begin{align}
    \left[\begin{matrix} P_1^{n}  \\ P_2^{n} \\ P_3^{n} \\ \vdots \\ P_M^{n} \end{matrix} \right]  = 
    \left[\begin{matrix} P_1  \\ P_2 \\ P_3 \\ \vdots \\ P_M \end{matrix} \right] +
    \left[\begin{matrix}
            \partial_{11} & \partial_{12} &  \partial_{14} \\
            \partial_{21} & \partial_{22} &  \partial_{24} \\
            \partial_{31} & \partial_{32} &  \partial_{34} \\
            \vdots & \vdots &  \vdots \\
            \partial_{M1} & \partial_{M2} &  \partial_{M4}
        \end{matrix} \right] \cdot 
        \left[\begin{matrix}  \left(D_1^n- D_1\right)  \\ \left(D_2^n- D_2\right) \\ \left(D_4^n-D_4\right)  \end{matrix} \right] 
        \label{eq5}
\end{align}
}

\subsubsection{Incremental Approaches:} Incremental updates calculate parity deltas between the updated data and corresponding original data without involving other unmodified data blocks. The incremental update method is better suited for small-grained update scenarios as it avoids introducing significant random-access and network overhead. It can be optimized using the mathematical principles of the updating formula, as shown in Equation (\ref{eq1}-\ref{eq5}). 

When a data block $(D_1)$ in a stripe is updated, the incremental update mechanism first computes the data delta $(D_1^n - D_1^{n-1})$ between the old and new data. Then, the update process generates a corresponding parity delta for each parity block, according to the Equation (\ref{eq2}). As shown in Equation (\ref{eq2}), $ P_1^n $ represents the calculated parity data, $ P_1^{n-1} $ denotes the original parity data, $(D_1^n-D_1^{n-1})$ denotes the data delta, and $\partial_{11}\ast(D_1^n-D_1^{n-1})$ denotes the parity delta for $ P_1 $. For a data block that undergoes $n$ updates, Equation (\ref{eq3}) is used to calculate its corresponding parity block. By applying the associative law, Equation (\ref{eq3}) is transformed into Equation (\ref{eq4}), showing that the latest update for the same location is considered valid, thus avoiding random-access, network, and calculation overheads from repeated updates. Merging multiple updates of the same address across data blocks in the same stripe into one parity delta significantly reduces network traffic, as shown in Equation (\ref{eq5}) and demonstrated in CoRD \cite{ref55}.

\subsection{State-of-the-Art Erasure Coding Update Methods}
\label{relate works}
\begin{figure}
    \centering
    \includegraphics[width=\linewidth]{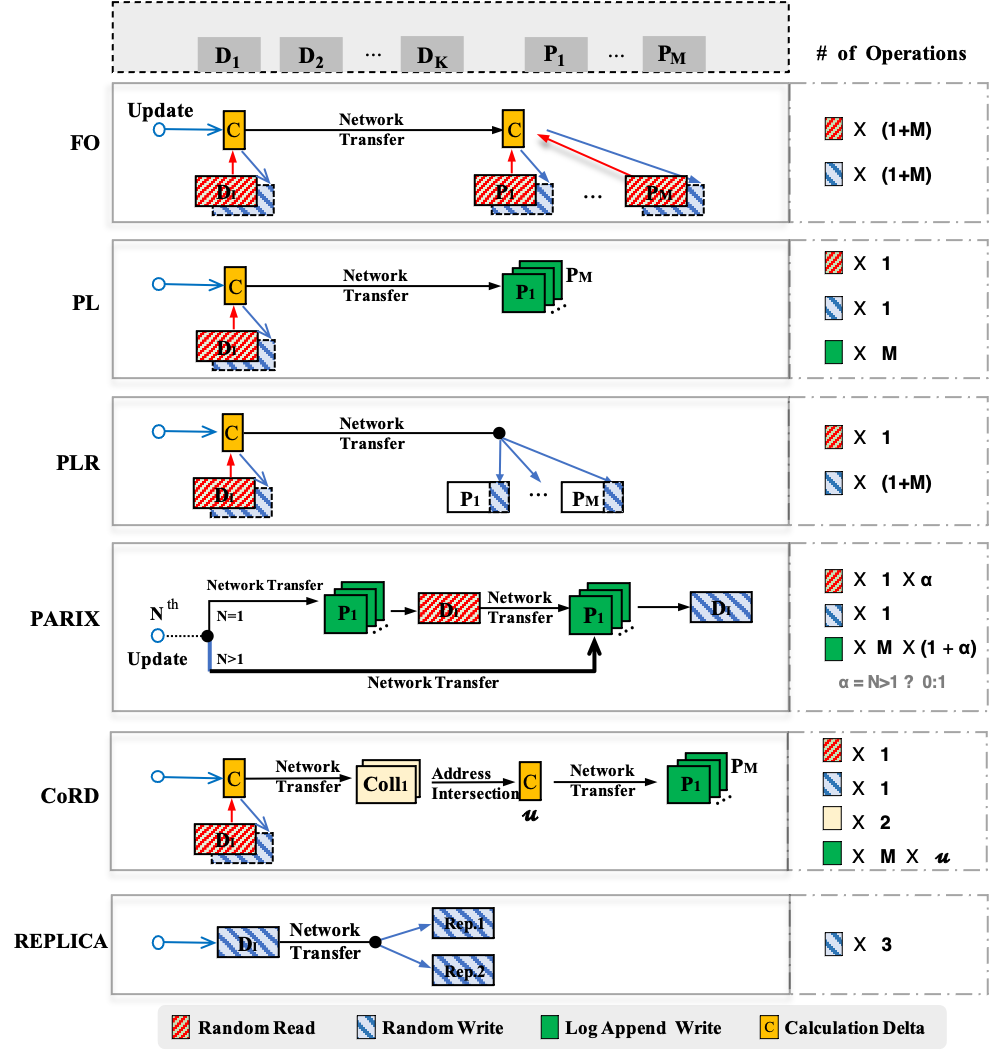}
    \caption{The update process of erasure coding and replica. Note that the operation on $ P_1 $ is also performed on $ P_2 $ - $ P_M $ although it is not illustrated.}
    \label{fig:F1}
\end{figure}

We illustrate the incremental update mechanism of erasure codes and the update mechanism of replicas in Fig. \ref{fig:F1}. The vertical axis represents the different mechanisms, while the horizontal axis represents the execution time, i.e., the update latency. It illustrates that FO exhibits the largest update latency, whereas the replica mechanism has the lowest update latency.

The Full-Overwrite (FO) \cite{ref58} method requires in-place updates for both data and parity blocks. The FO mode has the longest update path among all update methods, necessitating sequential updates to both data and parity blocks. During the update process, all operations are small-grained and random. For SSDs, small-grained random operations exhibit significantly higher latency and lower throughput compared to sequential operations, resulting in a performance gap several times greater. As depicted in Fig. \ref{fig:F1}, the random overhead of FO is the highest, and its update process is the most time-consuming.

The full-logging (FL) adopts logs to append the data/parity delta to avoid the disk read overhead of the old block for frequent updates. FL is widely used in enterprise storage systems such as Azure \cite{ref11} and GFS \cite{ref1}. However, the log spaces will merge with the old data when read requests come, it adversely affects data read performance. More importantly, FL demands a substantial amount of storage space for log data, while simultaneously incurring additional overheads associated with indexing, recycling, and reading the stored data. This contradicts the primary objective of introducing the erasure code mechanism, which is to reduce storage overhead. In addition, FL employs a single log structure, which renders the operations of log reading, appending, and recycling mutually exclusive. During the log recycling process, the log becomes inaccessible, thereby disrupting data reading and appending operations. Therefore, in the process of log design, it is essential to avoid the creation of large-capacity, highly mutually exclusive, and long-term storage logs.

The Parity Logging method (PL) \cite{ref6, ref7, ref8} effectively eliminates the overhead associated with in-place updates of parity blocks by using a parity-logging mechanism. In PL, updates to data blocks are performed in-place, it needs to perform a time-consuming write-after-read process to calculate the parity delta,  which is forwarded to parity logs corresponding to each parity block in the same stripe, leading to long latency, as shown in Fig. \ref{fig:F1}, the introduce of parity log avoid the random-access overhead for parity blocks. The log recycling process also involves lots of random access, which leads to poor log recycle efficiency and affects system reliability.

The Parity-Logging with Reserved Space (PLR) method \cite{ref2} utilizes dedicated log space adjacent to the parity block for storing parity deltas, thereby reducing the random-access overhead associated with reading parity deltas during the recycling process compared to PL, resulting in higher recycle efficiency. However, similar to PL, PLR requires an time-consuming write-after-read process for computing parity deltas. The distribution of log spaces adjacent to parity blocks across different locations of the storage device leads to random access during the appending operation to the parity log of numerous parity blocks, potentially resulting in disk space fragmentation due to frequent log appending and recycling.

The Partial Write (PARIX) method \cite{ref3, ref59} directly forwards update requests to the parity log, bypassing the expensive parity delta calculation via write-after-read. However, if a location is updated once $(N=1)$, the original data must be read and forwarded to the parity log separately, resulting in 2$\times$ network latency, as depicted in Fig. \ref{fig:F1}. PARIX stores both updated and original data for parity delta calculation in the parity log, rather than the parity deltas themselves. Among mainstream erasure code update schemes, only PARIX considers the temporal locality of data updates, effectively reducing the random read overhead 
from multiple updates at the same location. However, it does not fully exploit temporal locality to reduce random write overhead due to the in-place update mechanism and does not leverage spatial locality to reduce the impact of random access on disk performance.

The  Combination of Raid-based and Delta-based scheme (CoRD)  \cite{ref55} design a combination of raid-based and delta-based schemes to minimizes the update traffic. It selects the collector that collects and aggregates the deltas of updated parts to reduce update traffic, and flipping some dedicated blocks is proposed to enhance the effectiveness on suppressing the update traffic. Nevertheless, this approach necessitates a time-consuming write-after-read operation to compute the data delta. Moreover, CoRD overlooks parallelism and throughput considerations, leading to a single log becoming a critical bottleneck in the operational process, which significantly hampers update efficiency.  Additionally, similar to other methodologies, CoRD fails to address the localized optimization of data blocks, indicating substantial potential for further enhancement.

\subsection{Issues of Existing Methods and Research Challenges}
\label{challenges}

\subsubsection{High Update Latency}
The update latency remains high due to the lengthy data update path and the inherent nature of random access operations during the update process. \textbf{1) The long update path:} The incremental update mechanism for erasure codes requires updates to both data blocks and multiple corresponding parity blocks along the update path. Although the parity log postpones the update procedure of parity blocks, it still necessitates in-place updates of data blocks to calculate the data delta. This delta must then be forwarded and appended to the corresponding parity logs for the parity blocks within the same stripe. As a result, the potential for optimizing update latency is limited by the in-place updates of data blocks, which involve a time-consuming read-write process to compute the data delta. \textbf{2) The random-access latency:} The read-write random access latency of SSDs remains several times higher than that of sequential read-write access pattern, particularly under heavy IO loads, where the random access latency can increase to hundreds of microseconds.

To reduce update latency, it is crucial to shorten the erasure code update path and minimize random access costs. As shown in Fig. \ref{fig:F1}, the replica mechanism has the lowest update latency, particularly for sequential appending operations to copies. Thus, we propose transforming the in-place update mode of data blocks by changing the critical path of erasure code updates to a sequential append process based on the replica mechanism, thereby achieving shorter update latency. This approach ensures reliability while avoiding time-consuming read-write operations for compute data deltas.

\subsubsection{Consistency Issues accused by ParityLogs}
The delayed recycling of parity logs negatively impacts data consistency during data recovery. Most incremental update methods defer the recycling of parity logs until storage space reaches a certain threshold or data loss occurs. In the event of a disk or node failure, the system must recycle these logs before initiating data recovery. Prolonged log recycling may lead to secondary data loss, while log loss can compromise data integrity, both contributing to data consistency issues. Therefore, log items must be merged into the original data blocks and corresponding parity blocks before the actual recovery process begins. To ensure data consistency, a timely log recycling strategy should be incorporated into the design of the update solution.

\subsubsection{Low Throughput}
High throughput is essential for handling burst data and high-performance applications. The current incremental update mechanism typically employs the parity log as a boundary point, dividing the update process into two phases: the synchronous phase and the asynchronous phase. During these phases, numerous concurrent random fine-grained accesses prevent optimal storage device performance. In the synchronous phase, the update process performs in-place updates to calculate the data delta, which is then appended to the parity logs associated with the corresponding parity blocks in the same stripe. This stage involves two small-grained random accesses—read and write—which are key bottlenecks for improving update throughput. In the asynchronous phase, the update process merges multiple data deltas into the parity blocks, requiring two additional small-grained random-access operations, further affecting performance.

Data access patterns frequently exhibit spatio-temporal locality, which can be exploited to significantly reduce the number of actual accesses. There is a substantial performance gap between random and sequential access on SSD devices in terms of IOPS, read/write latency, and aggregate bandwidth. At the algorithmic software level, beyond optimizing the order of execution or reducing the frequency of accesses, there are limited methods to efficiently enhance small-grained random access on storage devices. Analysis of the Ten-Cloud Trace \cite{ref49} indicates that over 80\% of datasets processed less than 5\% of their total data volume, while more than 10\% of datasets processed less than 0.5\% of their data. By leveraging spatio-temporal locality, the log recycle process can reduce random access operations and associated computational and network overhead by one to two orders of magnitude. Consequently, optimizing erasure code updates necessitates a thorough investigation into the spatio-temporal characteristics of the access process.

\subsubsection{Low Lifespan}
The erasure code update operation incurs a write penalty, causing overwrite on data blocks and multiple parity blocks. frequent small-grained overwrites cause lots of erase operations on flash cells. Since SSD flash cells have a limited number of erase cycles \cite{ref56}, these overwrites shorten SSD lifespan, despite wear-leveling algorithms. Therefore, when designing the update algorithm, both write amplification and lifespan concerns must be considered. As noted earlier, leveraging spatio-temporal locality to reduce update requests and increase access granularity can significantly enhance the SSD lifespan.

\section{Design of Proposed TSUE}

\begin{figure*}
    \centering 
    \includegraphics[width=\linewidth]{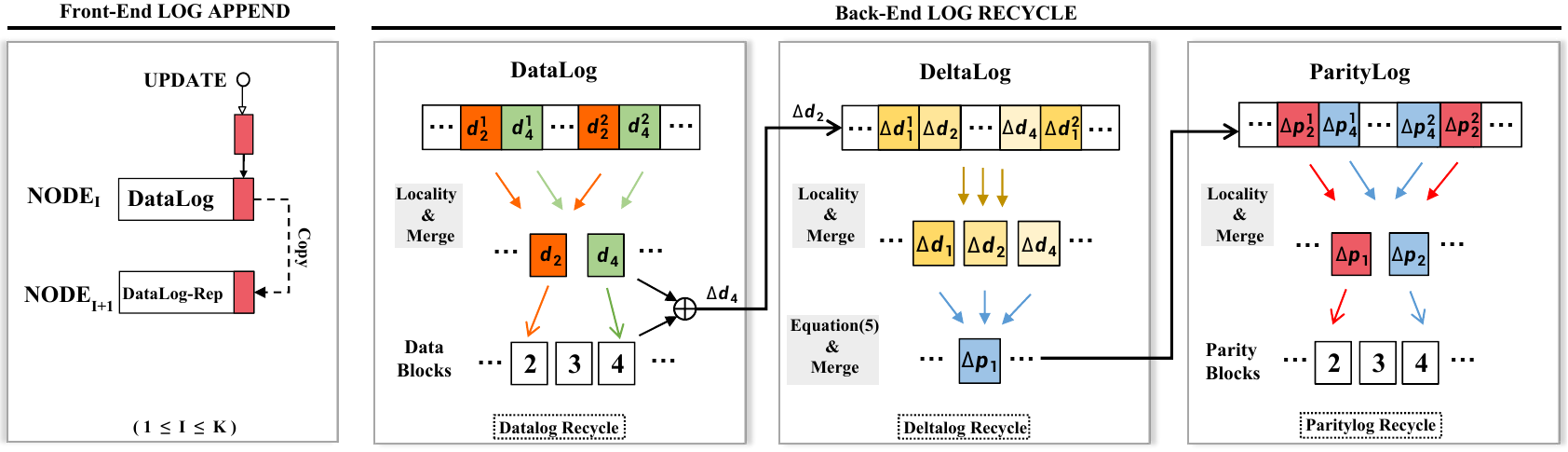}
    \caption{The workflow of TSUE. There is only 2 copies for DataLog in SSD cluster, and 3 copies for DataLog in HDD cluster. \newline
    \textbf{Note:} Due to the varying performance of different devices, the retention time of log items in memory differs. For SSDs, which have a lifecycle of approximately 10 seconds (Table. \ref{table2}), 2 copies are sufficient to ensure data reliability. In contrast, for HDDs, the number of log copies is designed to be 3.}
    \label{fig:F2}
\end{figure*}

\subsection{Overview of TSUE Architecture}
\label{overall workflow}

To address the aforementioned requirements, we propose a two-stage erasure code update method based on a three-layer log structure (TSUE). The design and implementation of TSUE addresses several issues in the update process, achieving the following goals:

\textbf{Low latency:} TSUE introduces a data log to record the received updated data at the data block side, thereby avoiding the I/O overhead and access latency associated with in-place updates of data blocks in state-of-the-art methods. The erasure code update process is divided by TSUE into two stages: synchronous front-end log appending operation and asynchronous back-end log recycling operation, with the data log serving as the boundary point. During the synchronous update phase, erasure coding transforms the in-place update of data blocks into a sequential appending process of two-copy data logs, thereby significantly reducing the update delay associated with erasure coding.

\textbf{Real-Time Log Recycle:} The real-time log recycling mechanism mitigates potential impacts on data consistency and minimizes the effect of logging on the overall system's read performance.   

\textbf{High throughput:} TSUE leverages the spatio-temporal characteristics of access patterns and utilizes the features of original data, data deltas, and parity deltas to progressively reduce the amount of log recycling and network traffic in three-layer logs. TSUE has designed a log pool structure based on a FIFO queue to manage these logs. This structure is deployed in memory and persisted in SSD, enabling real-time log recycling and ensuring concurrent rapid-access performance during both log appending and recycling procedures. Additionally, TSUE imposes a quota on log memory consumption to effectively control log memory usage. During the log recycling process, TSUE significantly reduces the number of logs by merging update requests and intermediate data, thereby greatly improving data throughput.

\textbf{High lifespan:} The lifespan is significantly improved by the substantial reduction in overwrite operations caused by updates. TSUE significantly reduces the number of overwrite operations on data blocks and parity blocks, thereby decreasing the frequency of flash cells overwrites and extending the lifespan of SSDs.

TSUE divides the erasure code update process into two stages: synchronous front-end log appending and asynchronous back-end log recycling, with the data log serving as the boundary point. 
 
\subsubsection{Front-End of Log Appending}

During the log append stage, TSUE directly appends updates to the data log rather than using an in-place update mechanism to modify data blocks. When a node receives an update request, it appends the request to its own data log, forwards the request to the node hosting the corresponding replica of the data log, and returns an acknowledgment to clients indicating the completion of the update.

TSUE deploys data logging on the node where the corresponding data blocks are located, ensuring that all log items stored pertain exclusively to those specific data blocks. The updated requests to the same data block from multiple clients can be merged for processing, thereby fully exploiting spatial-temporal locality within the log. Once a predetermined size is reached by a given data log, it is marked as \texttt{RECYCABLE} and replaced with a new log unit in the log pool structure. At any given time, only one log unit in the data log pool is active, and the active log can receive update requests under read-write lock protection. 

In order to achieve high bandwidth, low latency, and optimal reliability, TSUE stores logs in both memory and storage devices to prevent data loss following power outages. To fully leverage the concurrent capabilities of SSDS, TSUE typically configures multiple log pools for each SSD device. This deployment enhances the concurrent read and write performance of SSDS. 

\subsubsection{Back-End Log Recycling}

The introduction of data logs separates the calculation and random-access processes from the synchronous stage to a background asynchronous stage. It moves the expensive write-after-read process to the asynchronous stage, reducing update delays during synchronization, as shown in Fig. \ref{fig:F2}.

To guarantee reliability, TSUE uses a real-time reclaim mechanism to ensure that log entries in the data log are promptly merged into their respective data blocks and parity blocks. If a node fails before recycling, the replica log allows the log to be read. Lost log data can be reloaded from the local node’s persistent storage or the replica log, after which the log can be recycled. Furthermore, if all replicas fail, data can be recovered from other blocks in the same stripe. The data log entries are asynchronously recycled in time to update both data and parity blocks, ensuring efficient updates and data reliability, with no negative impact on data recovery.

To reduce the overhead of single-level data log recycling, TSUE uses a three-level log structure consisting of data logs, delta logs, and parity logs. By leveraging the characteristics of data stored at each level and the spatio-temporal locality of access patterns, this structure enables the merging and concatenation of log data based on log level. Logs are progressively recycled, thereby distributing the random-access, computation, and network overheads.

\textbf{ Data Log Recycle:} During the data log recycling process, TSUE performs read-write operations to calculate data deltas and overwrite original data blocks. A two-level index is employed to efficiently manage log items based on data blocks and intra-block offsets. Duplicate and adjacent data are reorganized according to spatio-temporal access patterns. This consolidation of numerous small-sized random accesses into fewer larger-sized accesses within the log facilitates more efficient data block recycling. During data log recycling, the original data is read, and the data delta between the original and updated data is calculated. This data delta is then forwarded and appended to the delta log of the OSD hosting the first and second parity blocks in the same stripe. 

\textbf{ Delta Log Recycle:} During recycling, data deltas are merged and combined based on spatio-temporal locality without actual data reads or writes. Deltas for the same position are merged using an XOR operation, as described in Equation (\ref{eq3}), while adjacent deltas are concatenated. Exploiting spatial locality, deltas for the same address across different data blocks in the same stripe are merged into a single parity delta, as described in Equation (\ref{eq5}). This approach enables the delta log to significantly reduce I/O operations and network traffic compared to the data log.

\textbf{ Parity Log Recycle:} The append process of the parity log is similar to that of the delta log. During parity log recycling, parity deltas undergo XOR operations with the corresponding parity block to generate new parity data, which directly overwrites the original parity block. The presence of the delta log simplifies parity updates by reducing matrix multiplication operations to a single XOR operation, thereby lowering computational complexity and significantly reducing the number of parity deltas appended to the parity log. During parity log recycling, adjacent parity deltas are combined, and parity deltas at the same location are merged using XOR operations based on spatio-temporal locality. This approach minimizes disk I/O overhead and enhances system performance during the recycling process.

\subsection{Log Pool Structure of TSUE}
\label{log pool}

\begin{comment}
   \textcolor{blue}{Multiple update records within the DataLog index are consolidated into one index with unique and large granularity. Multiple update records for the same and adjacent locations in the DeltaLog and ParityLog Index, delineated by a dashed gray line, are merged into one larger-grained read/write operation.} 
\end{comment}

TSUE designs a log pool structure \cite{ref17} based on a FIFO queue to efficiently manage the three-level log architecture and concurrently support log writes and reads, which enhances both append and recycle performance. The main contributions of the log pool structure are as follows:  Managing log units via a FIFO queue, which supports concurrent data appending and recycling. Utilizing the built-in two-level index structure within log units reduces the volume of log recycling by leveraging spatio-temporal locality of access patterns.  Dynamically expanding and contracting the log pool based on access hotspots. Retaining the internal data and indexes of recycled log units until they are reused as active units for new data additions, allowing the log unit to function as a read cache and enhancing data read performance. According to the characteristics of storage devices like HDDs and SSDs, 1-N log pools can be configured per device to support concurrent access capabilities.

\subsubsection{FIFO-based Log Pool Structure}

\begin{figure}
    \centering
    \includegraphics[width=\linewidth]{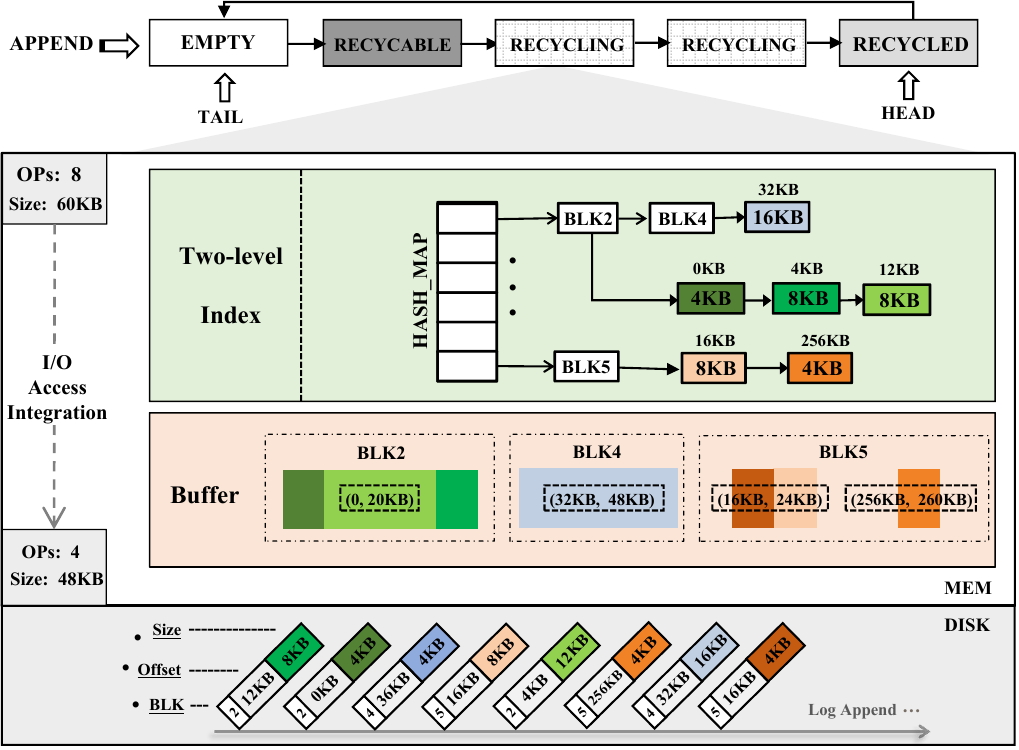}
    \caption{Log pool structure of TSUE.}
%    \vspace{-2mm}
    \label{fig:F3}
\end{figure}

TSUE has developed a log pool structure based on the FIFO queue to ensure high performance for both log appending and recycling simultaneously, as depicted in Fig. \ref{fig:F3}. To enhance read-write performance, the log pool structure comprises two components: memory and disk. The log unit index and log unit buffer in memory serve as a cache, with the memory buffer persisted to disk for reliability. To prevent TSUE from consuming excessive memory, a memory usage limit is imposed, which necessitates sufficiently fast recycling efficiency to ensure log appending performance.

The log unit in TSUE has four states: EMPTY, RECYCLABLE, RECYCLING, RECYCLED, as shown in Fig. \ref{fig:F3}. Initially, all data logs are marked as EMPTY. At any given time, only one active log unit marked as EMPTY is available for data appending. Once the active log unit is filled up, it is marked as RECYCLABLE, indicating that it can be recycled. Multiple log units marked as RECYCLABLE can be recycled concurrently with their own independent indexes. Once the log unit marked as RECYCABLE is attach to recycle thread to perform recycling, it is remarked as RECYCLING. After complete recycling, it is then marked as RECYCLED.  The active log unit resides at the end of the log queue. If there are no empty log units in the current pool, the next available log unit marked as RECYCLED is remarked as EMPTY, and its remaining index entries are freed to accept new data appending.  A log unit can serve as a read cache to accelerate read performance until it is re-marked as EMPTY.

Timely recycling is crucial for supporting high-speed log appending operations within limited space. To mitigate the adverse impact of log recycling on data recovery and minimize log space occupation in both memory and disk, TSUE implements a real-time recycling strategy for all logs, ensuring prompt merging into data blocks and parity blocks. 

In the three-layer log architecture, TSUE maintains multiple log units in a FIFO queue, enabling concurrent block-level recycling of these units to enhance both log appending and recycling performance. TSUE constructs a recycling thread pool where log entries in log units are assigned to multiple recycling threads on a per-block basis, allowing multiple recyclable log units to be processed concurrently. Log records for the same block are assigned to the same recycle thread to maintain consistency. When multiple log entries belonging to the same block are distributed across concurrently recycling log units during parallel recycling, they are sequentially merged and aggregated according to their order.

On the OSD node, the OSD determines the corresponding log pool in the data log, delta log, or parity log based on the hash value of the file identifier, which comprises the inode number, stripe number, and block number. When the active log unit in log pool on the OSD becomes full, the system switches to the next available log unit in the log pool, as illustrated in Fig. \ref{fig:F3}. The filled active log unit is marked as RECYCLABLE. Log units marked as RECYCLABLE are recycled in blocks by different recycling threads. Once all blocks of a log unit have been recycled, it is then remarked as RECYCLED.

\subsubsection{Scalable Updating Hotspot}

The TSUE log structure is adaptive and scalable, optimizing log space utilization and update efficiency across various scenarios. The scalable log pool can adjust its size based on load, particularly addressing memory constraints. The log structure consists of fixed-size log units, with the flexibility to adjust the number of logs according to workload demands, ensuring scalability.  In high random-access scenarios, TSUE expands its log space for recycling, while in low-overhead scenarios, it maintains minimal space for fast log access during append and recycle. Unused log space is released when updates are absent or data is quickly recycled. Log units are organized in a FIFO queue, allowing unused units to be removed and new ones to be added. TSUE reduces lock protection domains by assigning independent index for each log unit and assign 1-N log pools per device to support concurrent access capabilities, ensuring efficient I/O concurrency among disks within the node.

\subsection{Spatial-Temporal Locality of Access Pattern}
\begin{comment}
\begin{figure}
    \centering
    \includegraphics[width=\linewidth]{List-Update/F4-update-IPDPS.pdf}
    \caption{the utilization of spatial-temporal locality in TSUE.  The utilization of locality is based on a two-level hash index, blocks are organized by hash-map, and update records are organized by offset and size in belonged blocks. }
    \label{fig:F4}
\end{figure}
\end{comment}

TSUE leverages spatio-temporal locality to identify repeated and adjacent update requests, reducing random access volumes and improving access granularity, thus enhancing disk throughput. o further exploit the spatio-temporal locality of log records, we propose a two-level index structure that supports block and offset.
\begin{comment}
As depicted in Fig. \ref{fig:F4}, during the process of appending records to the data log, 8 update records totaling 60KB are appended. By leveraging spatio-temporal locality mining, these update records are aggregated into 4 large-grained records totaling 48KB during the log recycling process, resulting in an increase in average access granularity from 7.5KB to 12KB.
\end{comment}

\subsubsection{High-speed Two-level Indexing}

To fully exploit the spatio-temporal locality of log access patterns and minimize the data processed during recycle, we design a two-level index structure that classifies data by blocks and sorts by offset, as shown in Fig. \ref{fig:F3}. The first level is a block index, where data blocks are organized in a hash table for efficient retrieval. The second level is offset and size index, where items are sorted in a linked list by offset and merged based on size. The positions of sequentially appended update records in both data and parity blocks are random. By re-indexing the triples $<$block, offset, size$>$, all update records are organized in blocks and sorted by offset and size. Merge and aggregation operations for update requests are executed based on this indexing mechanism. The repeated or adjacent records can be identified easily according on the index. To enhance query speed while avoiding lock conflicts, a bitmap is associated with the block-level index to quickly determine query hits and eliminate unnecessary linked list queries.

\subsubsection{Utilization of Spatial-temporal Locality}
\label{sptial-temporal locality}

The utilization of spatial-temporal locality \cite{ref9} can be leveraged to identify random accesses at the same and neighboring locations in update requests, thereby reducing request volume and enhancing access granularity. For data logs, the latest data can directly overwrite older data at the same location within data blocks, updates from adjacent locations can be consolidated into fewer, larger-sized log entries. For delta logs, each index entry carries information about its corresponding data block. Multiple data deltas for the same location across different data blocks within the same stripe can be merged and calculated into a parity delta according to Equation (\ref{eq5}). Multiple data deltas for the same location in the same data block can be calculated into one according to Equation (\ref{eq3}), and multiple adjacent data deltas can be merged into a larger one. For parity logs, multiple parity deltas for the same location can be consolidated into one parity delta, and multiple adjacent parity deltas can be merged into a larger one.

\subsubsection{Read Cache}
\label{read procedure}
The presence of a high-speed index allows the log to function as a read cache. When a data read occurs, if the data is found in the log, it can speed up data access efficiency. If not found, the data is read from the corresponding disk as usual. When a client receives a user access request through an API call, it first identifies the block's location based on the metadata server, then queries the node (storage server). Upon receiving a read request, the storage server first checks the TSUE log structure for the existence of data. If found, it returns the data directly; otherwise, it retrieves the data from the storage device. The storage system is designed to prevent the provision of stale data.

\subsection{Data Consistency Solution}
\label{consistency}
While updating erasure codes, some consistency challenges arise. To tackle these issues, TSUE's design resolves them as follows.

\textbf{ Update consistency when multiple client nodes update the same data block:} In cases of simultaneous updates by multiple clients on the same block, their requests are directed to the single designated node to which the block belongs. TSUE appends these requests to the data log in an ordered manner (based on arrival or queuing) and returns them to clients in the order they were added, thereby guaranteeing access consistency.

\textbf{ Data consistency among multiple parity logs:} It can be observed that the parity deltas corresponding to the same parity block may appear in an unordered manner within one or more parity log units. When calculating the latest parity data, all relevant parity deltas for a given position must be processed, regardless of their original order, as shown in Equation (2). Therefore, as long as all parity deltas can be appended to the parity log, their specific sequence becomes inconsequential.

\section{Implementation Details}

\begin{figure}[t]
    \centering
    \includegraphics[width=\linewidth]{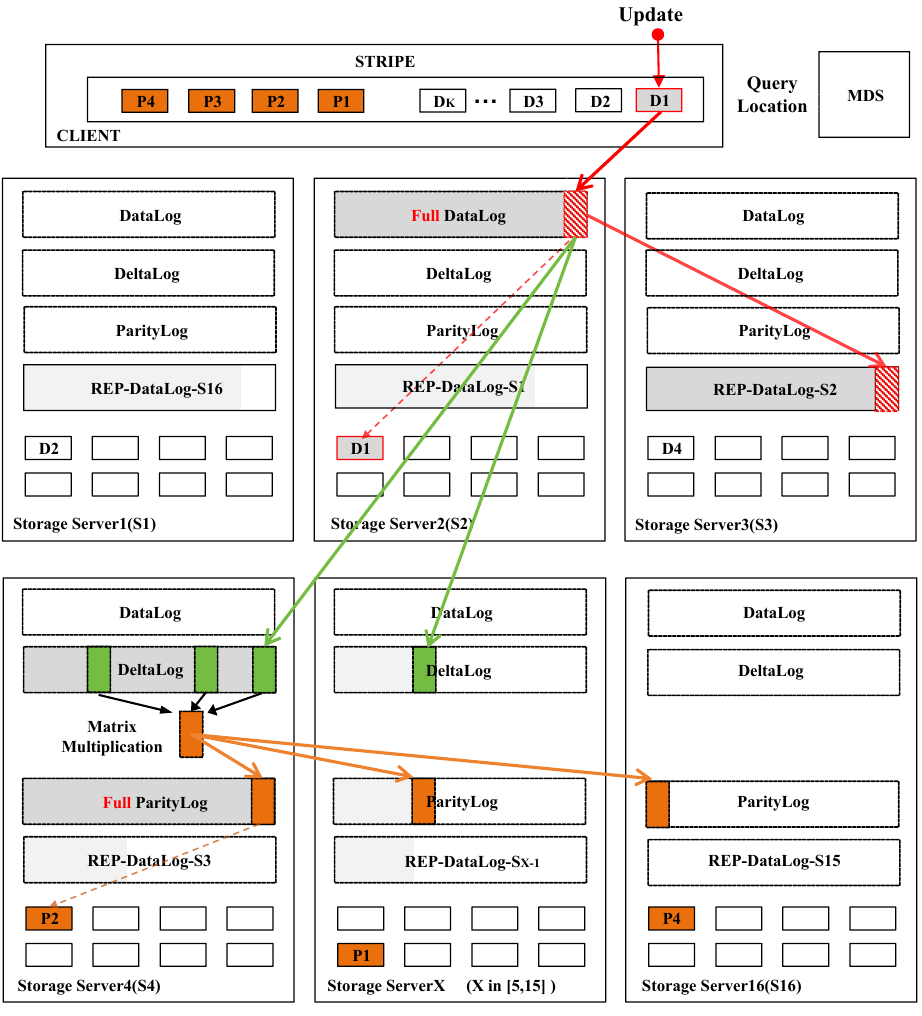}
    \caption{Architecture of ECFS.}
    \label{fig:F5}
\end{figure}

The TSUE technique is integrated into an self-developed erasure-coded distributed storage file system (ECFS) to enhance both log update and log recycle performance, while minimizing the overhead associated with file system updates. The ECFS is a distributed file system comprising of the metadata server (MDS), object storage device server (storage server, OSD), and CLIENT, as shown in Fig. \ref{fig:F5}. The CLIENT provides the POSIX access interface and handles the data encoding process, while the OSD stores and manages data blocks. The MDS manages the hierarchical tree structure of directories and files, tracks the locations of data blocks, and monitors the status of server nodes.

ECFS ensures data reliability by distributing both data blocks and parity blocks within each stripe across multiple OSDs. During system initialization, each OSD transmits its block information to MDS and regularly sends heartbeat messages to indicate its status. In case MDS fails to receive a heartbeat message from any OSD within a specific time interval, recovery tasks are initiated. The ECFS supports single-client exclusive write access to individual files while enabling concurrent update from multiple clients on a single large file through block-level locking mechanisms.

\textbf{Normal Write:} During the file writing process, the client partitions the file into fixed-size blocks according to the predefined block size. Through an encoding process, within each stripe, K data blocks generate M corresponding parity blocks. The client retrieves or queries the location information of the $K+M$ blocks within the stripe from the MDS, transmits each block to its corresponding OSD. The OSD receives and writes each respective block to its designated disk while updating its local block index to facilitate location queries during data reading.

\subsection{Log Organization}
\label{implementation}

Each OSD is equipped with a single SSD, four log pools are configured for each log structure, including data log, delta log and parity log, as shown in Fig. \ref{fig:F5}. Both the DataLog and the DeltaLog are maintained in duplicate. The replica of DataLog is stored solely in SSD, with no space allocated in memory. The copy of the data delta is stored in the DeltaLog corresponding to the second parity block within the stripe, while only the data delta associated with the first parity block within the stripe is recycled. TSUE determines the appropriate log pool in the DataLog, DeltaLog, or ParityLog based on the hash value of the file identifier, which includes the inode number, stripe number, and block number. This configuration ensures optimal read and write performance for logs on SSD devices while minimizing resource usage, particularly memory resources.

\subsection{Log Reliability}
In case of node failure, the data log on this node can be obtained from one of the nodes hosting its replica, and delta log entries on this node can be retrieved from the 2th delta log corresponding to each log entry. For the parity log, the parity block on the node becomes the most recent parity block following restoration. During recovery, the parity log content is already merged into the parity blocks, so it does not need to be restored.

\begin{comment}

In ECFS, data that is typically written undergoes encoding at the CLIENT and is subsequently distributed to the corresponding storage servers for storage. Updated data is directly transmitted to the storage server hosting the corresponding data block, then the update process is executed on both the corresponding data block and the parity check block. 
\end{comment}

\subsection{Update Procedure}
\label{implementaion:update procedure}

\textbf{Normal Write or Update Write:} 
The functional modules of TSUE are primarily implemented in the CLIENT and OSD components. The MDS constructs a scalable linked list for each file based on a page-level bitmap mapping structure, thereby distinguishing between write and update operations. In the CLIENT, the system differentiates between write and update operations using the linked list, and transmits both the updated data and corresponding markers to the designated OSD. Data that is typically written undergoes encoding at the CLIENT before being distributed to the corresponding storage servers for storage. Updated data, on the other hand, is directly transmitted to the storage server hosting the relevant data block, where the update process is executed on both the data block and its associated parity blocks.

\begin{figure*}[t]
    \centering

    \begin{subfigure}{0.24\linewidth}
        \centering
        \includegraphics[width=\linewidth]{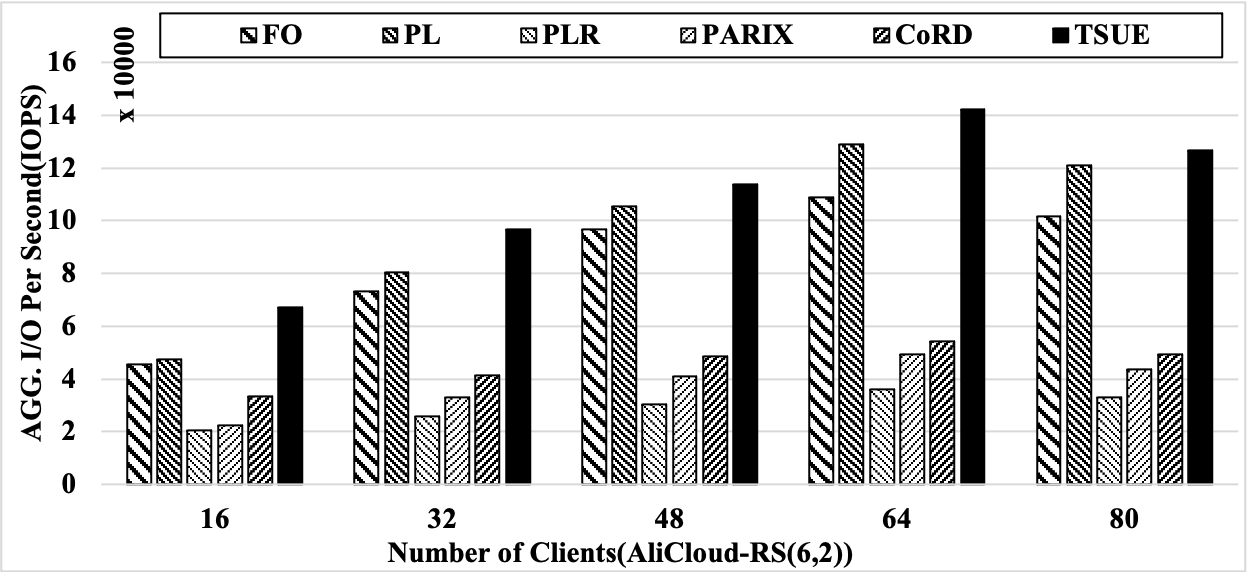}
        \caption{RS(6,2) and Ali-Cloud}
        \label{fig:F6-01}
    \end{subfigure}
    \begin{subfigure}{0.24\linewidth}
        \centering
        \includegraphics[width=\linewidth]{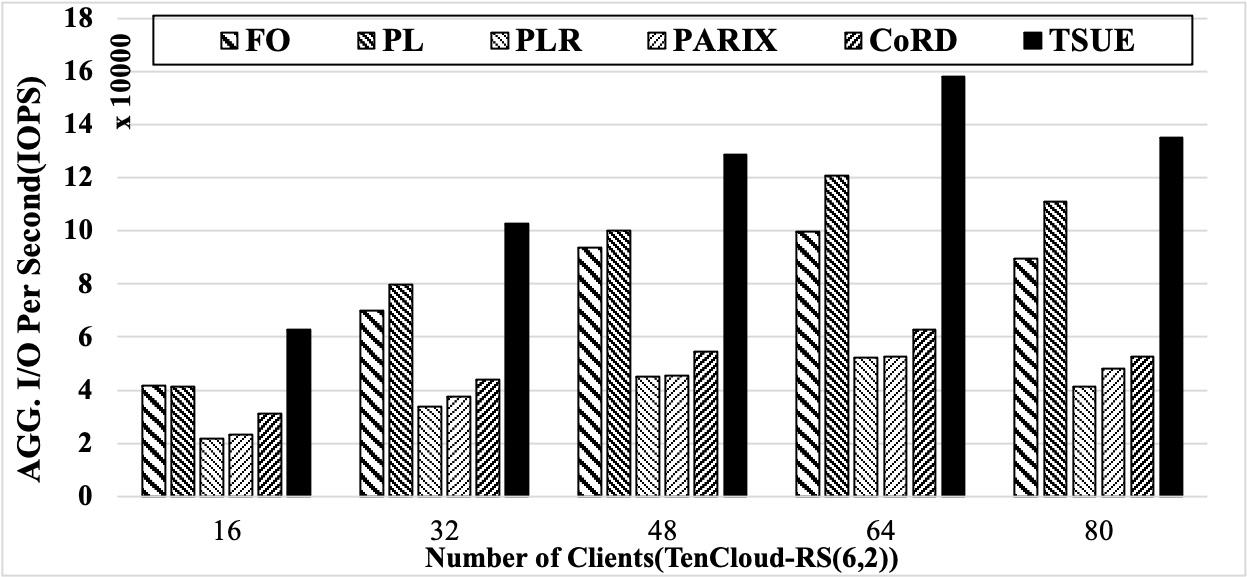}
        \caption{RS(6,2) and Ten-Cloud}
        \label{fig:F6-02}
    \end{subfigure}
    \begin{subfigure}{0.24\linewidth}
        \centering
        \includegraphics[width=\linewidth]{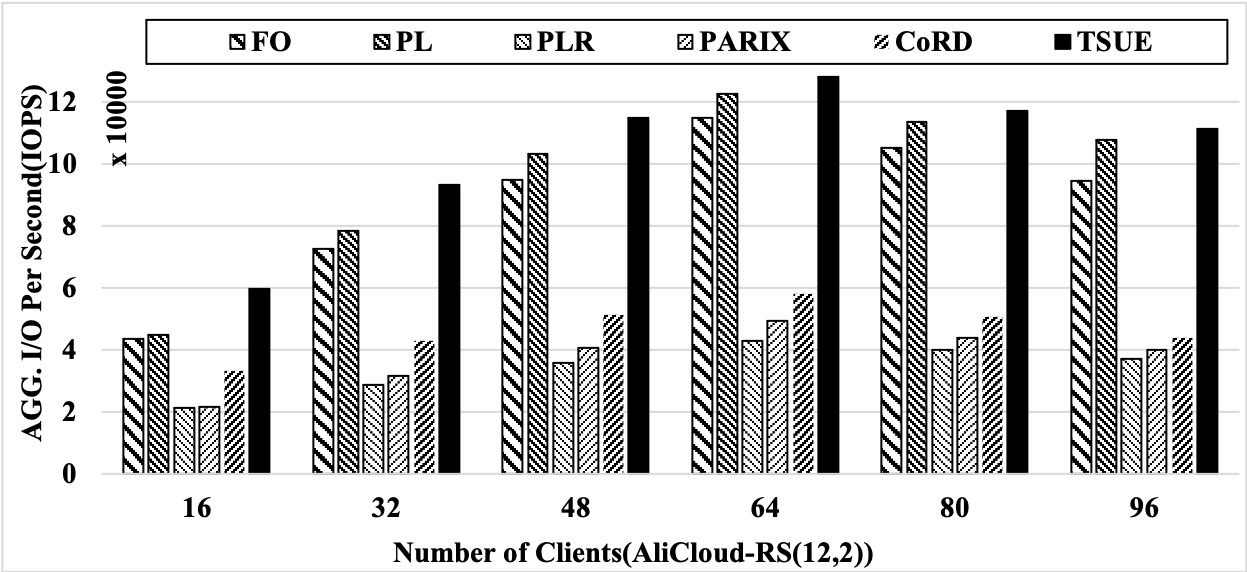}
        \caption{RS(12,2) and Ali-Cloud}
        \label{fig:F6-03}
    \end{subfigure}
    \begin{subfigure}{0.24\linewidth}
        \centering
        \includegraphics[width=\linewidth]{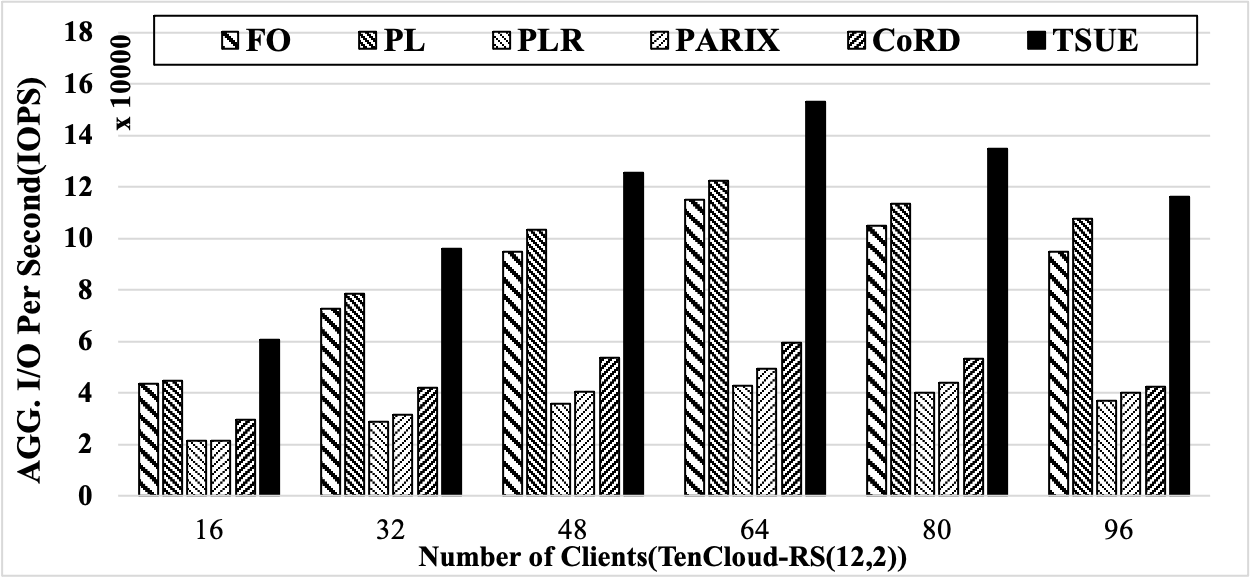}
        \caption{RS(12,2) and Ten-Cloud}
        \label{fig:F6-04}
    \end{subfigure}
    \begin{subfigure}{0.24\linewidth}
        \centering
        \includegraphics[width=\linewidth]{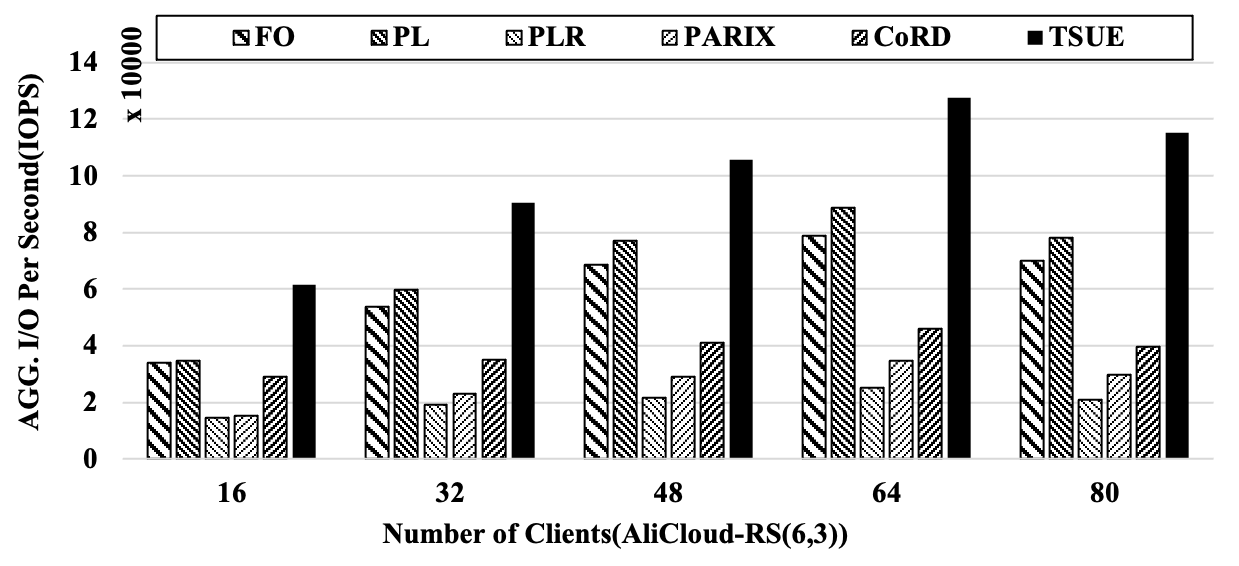}
        \caption{RS(6,3) and Ali-Cloud}
        \label{fig:F6-05}
    \end{subfigure}
    \begin{subfigure}{0.24\linewidth}
        \centering
        \includegraphics[width=\linewidth]{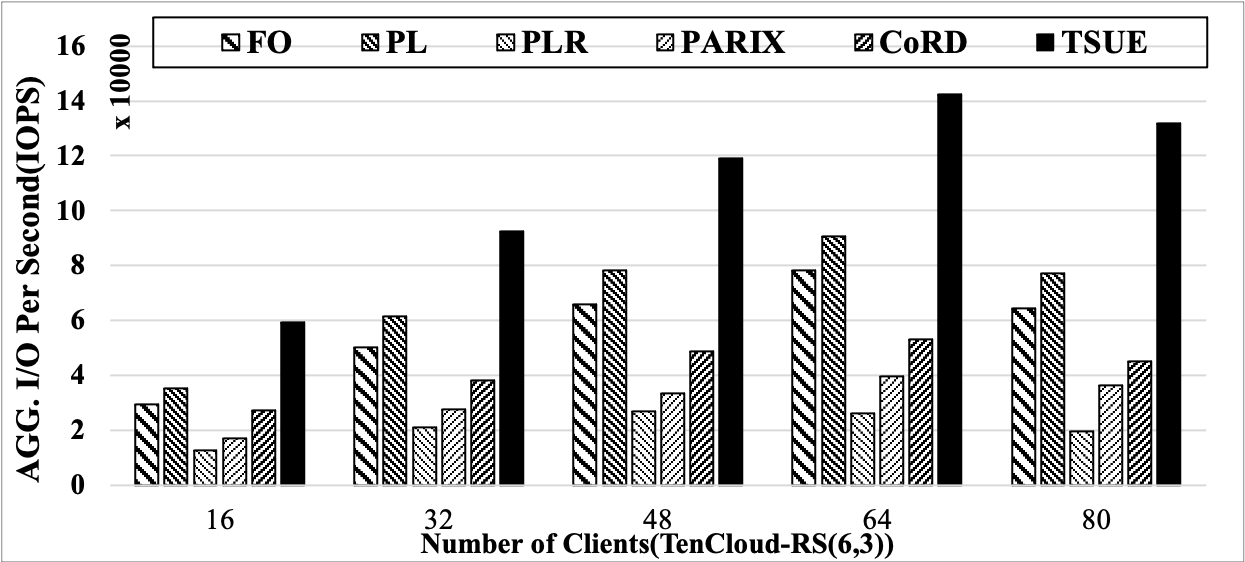}
        \caption{RS(6,3) and Ten-Cloud}
        \label{fig:F6-06}
    \end{subfigure}
    \begin{subfigure}{0.24\linewidth}
        \centering
        \includegraphics[width=\linewidth]{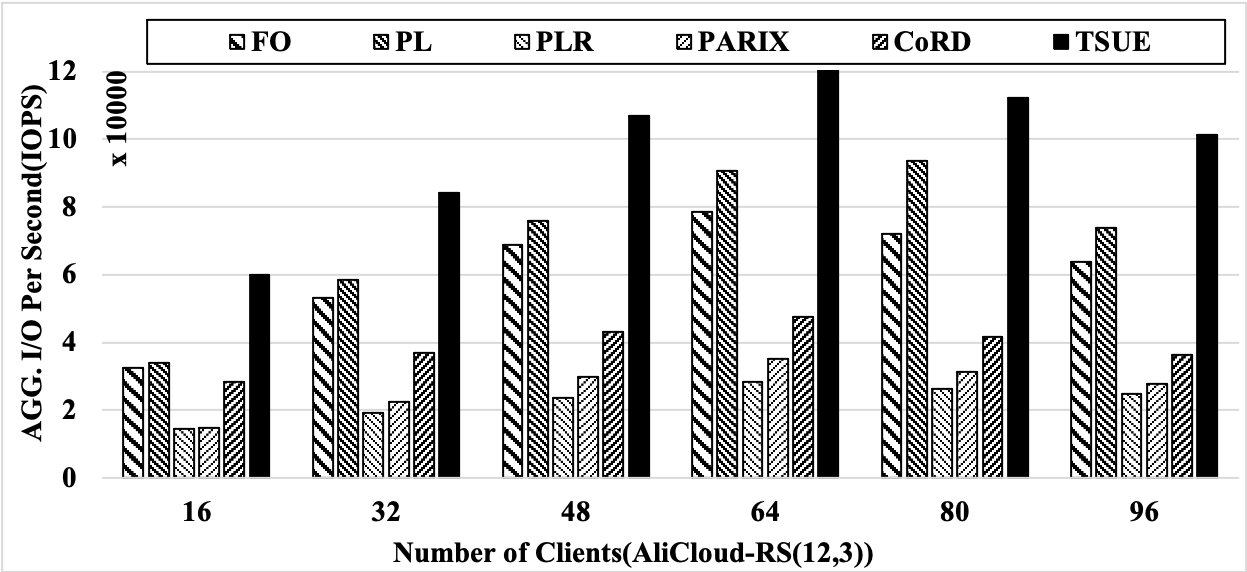}
        \caption{RS(12,3) and Ali-Cloud}
        \label{fig:F6-07}
    \end{subfigure}
    \begin{subfigure}{0.24\linewidth}
        \centering
        \includegraphics[width=\linewidth]{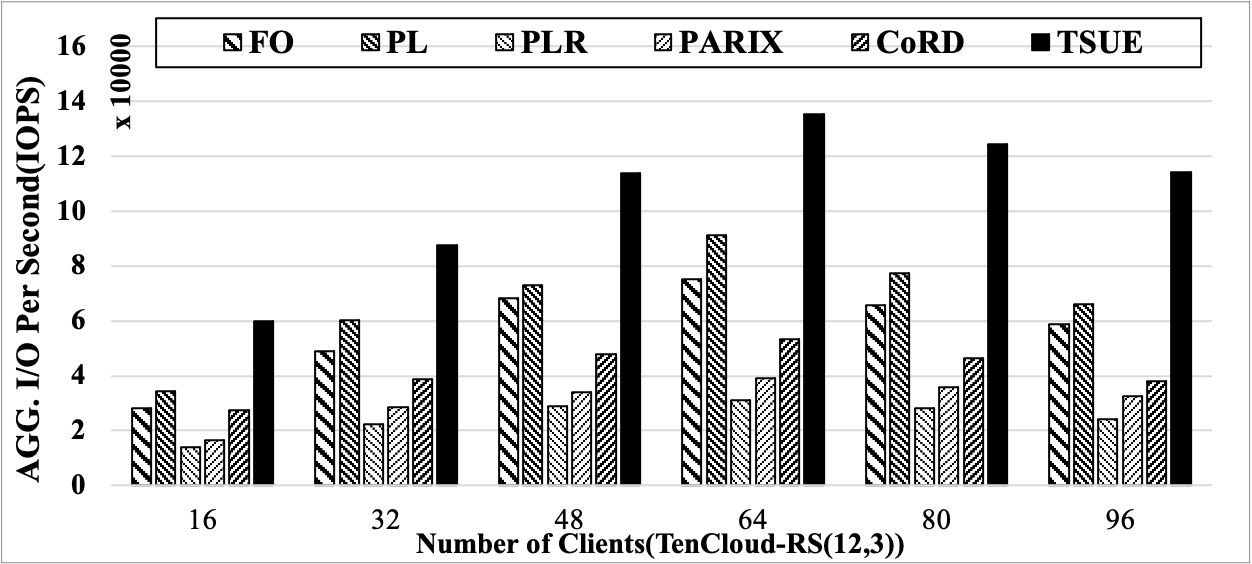}
        \caption{RS(12,3) and Ten-Cloud}
        \label{fig:F6-08}
    \end{subfigure}
    \begin{subfigure}{0.24\linewidth}
        \centering
        \includegraphics[width=\linewidth]{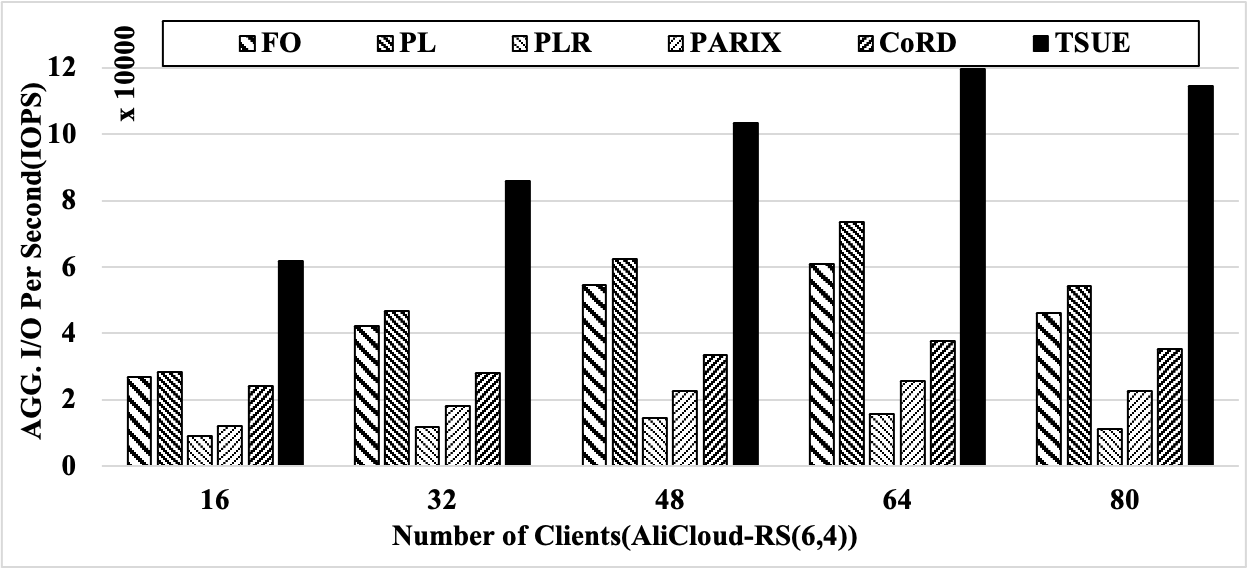}
        \caption{RS(6,4) and Ali-Cloud}
        \label{fig:F6-09}
    \end{subfigure}
    \begin{subfigure}{0.24\linewidth}
        \centering
        \includegraphics[width=\linewidth]{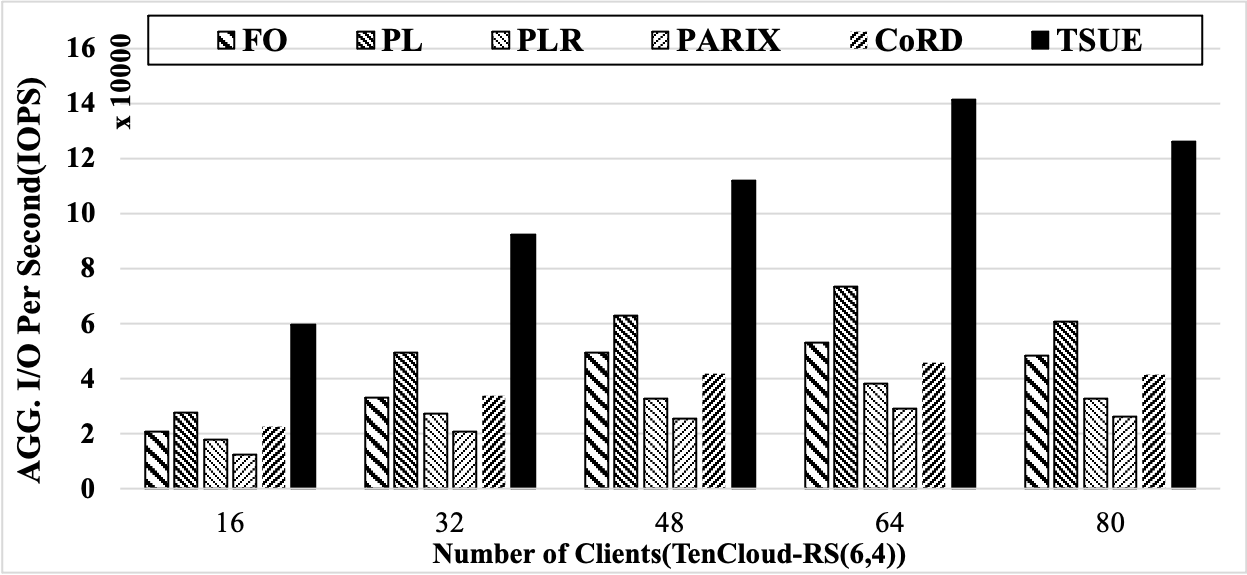}
        \caption{RS(6,4) and Ten-Cloud}
        \label{fig:F6-10}
    \end{subfigure}
    \begin{subfigure}{0.24\linewidth}
        \centering
        \includegraphics[width=\linewidth]{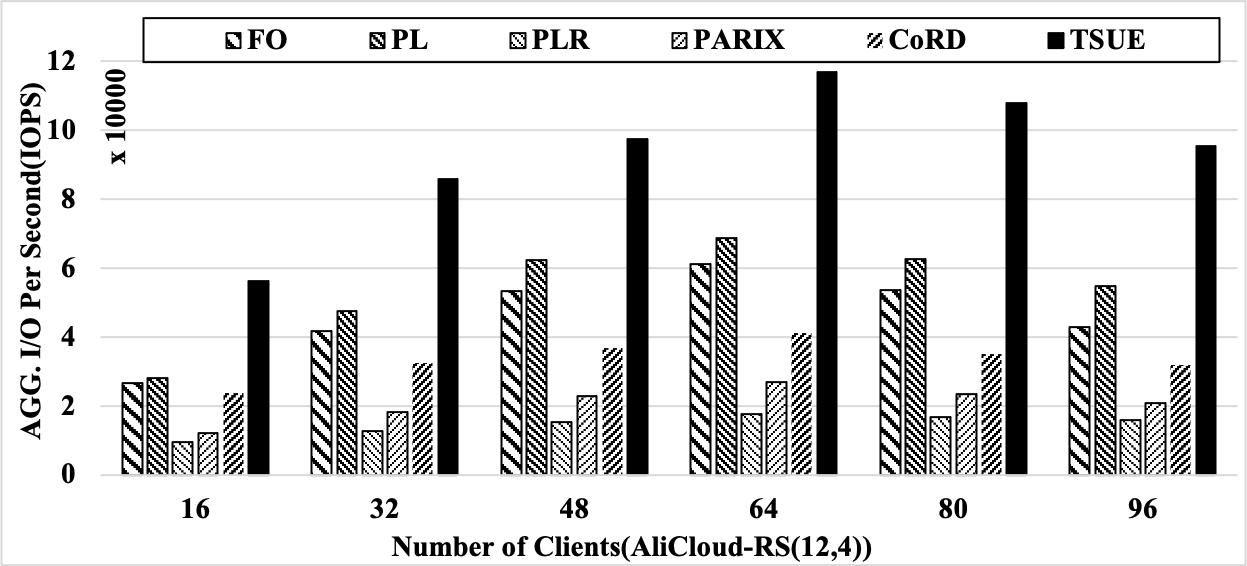}
        \caption{RS(12,4) and Ali-Cloud}
        \label{fig:F6-11}
    \end{subfigure}
    \begin{subfigure}{0.24\linewidth}
        \centering
        \includegraphics[width=\linewidth]{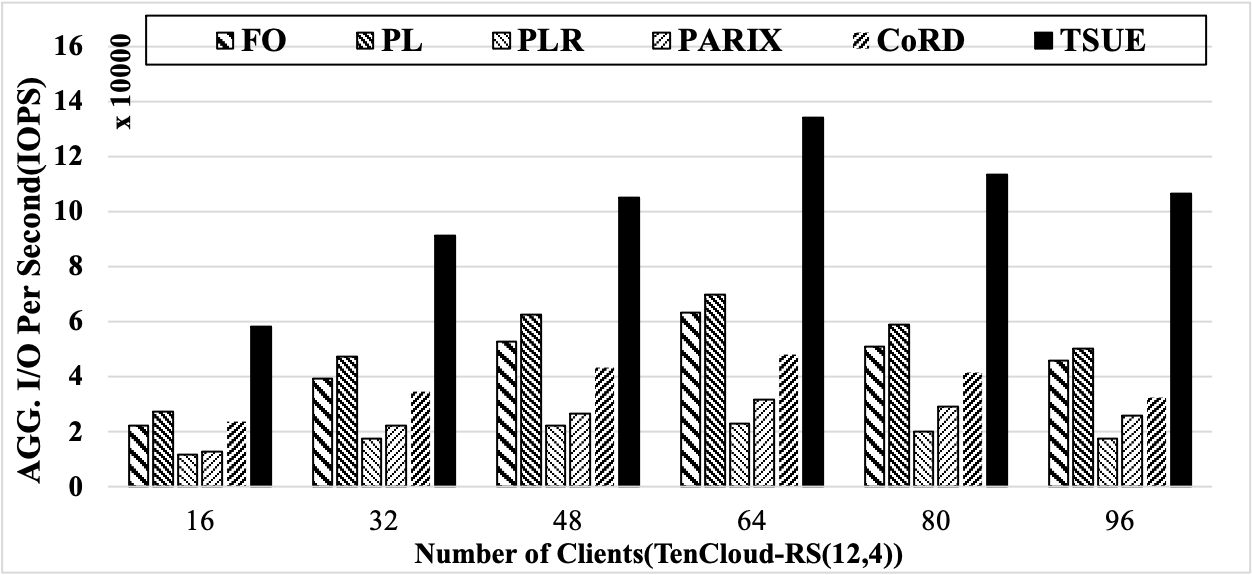}
        \caption{RS(12,4) and Ten-Cloud}
        \label{fig:F6-12}
    \end{subfigure}
    \caption{Performance evaluation of TSUE's update throughput with SSDs.}
    \label{fig:F6}
%    \vspace{-3mm}
\end{figure*}

\textbf{Front-End Update Procedure:} Upon receiving an update request from the Client, the OSD appends the request to the active log unit within the selected log pool in the data log, updates the log index, and forwards the request to the replica of this log pool in another OSD. Finally, the OSD sends a confirmation message to the Client, indicating that the update has been successfully completed.

\textbf{Back-End Recycle Procedure:}
TSUE uses a three-level log structure to organize log items. By exploiting the spatio-temporal locality of data access patterns, TSUE directly merges and concatenates original data in the DataLog. In the DeltaLog and ParityLog, it merges intermediate data at the same position according to Equations (\ref{eq3}, \ref{eq5}) and concatenates adjacent data, thereby reducing the volume of log recycling and improving recycling efficiency.

\section{Performance Evaluation} 

We implemented FO, PL, PLR, PARIX, CoRD, and TSUE within the ECFS file system. The implementation of these methods in ECFS is confined to the CLIENT side and the OSD side. On the CLIENT side, it is responsible for identifying the node where the updated data resides, distinguishing between regular writes and update writes, and transmitting the updated data along with update flags to the corresponding OSD server. On the OSD side, core mechanisms of the update process are implemented, including data reading, delta computation, writing back, log management, intermediate data forwarding, and log recycling processes among OSD servers.

We tested and analyzed their performance in an SSD environment using block-level traces from  Ali-Cloud \cite{ref48} and Ten-Cloud \cite{ref49}. Additionally, we evaluated their performance in an HDD environment using the MSR Cambridge Trace.

\subsection{Experimental Platform}
\label{experimental}
To validate the optimization effect of the TSUE mechanism on SSD Cluster, we configure a 16-nodes test-bed from chameleon cloud platform \cite{ref54}. Each node was equipped with 256GB memory and a 400GB SSD, interconnected via a 25Gb/s Ethernet switch. On this test-bed, we construct an ECFS filesystem with a capacity of 6.4TB.

\subsection{Evaluation on Real-Workloads}
\label{high-speed-test}
As depicted in the Fig. \ref{fig:F6}, we conducted various update tests on FO, PL, PLR, PARIX, CoRD and TSUE using Ali-Cloud \cite{ref48} and Ten-Cloud \cite{ref49} traces for 6 RS($K$, $M$) codes with diverse $K$ and $M$ parameters. 
\textbf{RS(6,2) and RS(12,2):} TSUE demonstrated performance levels reaching 1.5$\times$ that of FO, 1.5$\times$ that of PL, 3.9$\times$ that of PLR, 3.1$\times$ that of PARIX, and 2.6$\times$ that of CoRD.
\textbf{RS(6,3) and RS(12,3):} TSUE demonstrated performance levels reaching 2.1$\times$ that of FO, 1.8$\times$ that of PL, 6.7$\times$ that of PLR, 4.0$\times$ that of PARIX, and 3.0 $\times$ that of CoRD.
\textbf{RS(6,4) and RS(12,4):} TSUE demonstrated performance levels reaching 2.9$\times$ that of FO, 2.2$\times$ that of PL, 10.1$\times$ that of PLR, 5.1$\times$ that of PARIX, and 3.3$\times$ that of CoRD.

As can be seen from the above analysis, TSUE demonstrates the most outstanding performance overall. Specifically, TSUE exhibits superior performance under the Ten-Cloud trace compared to the Ali-Cloud trace. Additionally, TSUE's performance gradually improves with an increasing number of clients, and TSUE achieves optimal performance with 64 clients.

The performance advantage of TSUE becomes increasingly pronounced as the value of $M$ grows. When $M$ is small—note that its minimum value is 2, which has already been considered in this test—the performance improvement achieved by TSUE is relatively modest compared to other state-of-the-art (SOTA) mechanisms. This is because, with a small $M$, the overhead introduced by other mechanisms is also relatively low, particularly since the FO and PL mechanisms introduce minimal random access overhead. As $M$ increases, the random access overhead associated with these mechanisms becomes more significant. Despite this, TSUE consistently demonstrates effectiveness across all tested scenarios, with its optimization effect becoming progressively more substantial as $M$ grows.

CoRD is an update mechanism specifically designed to optimize network traffic. It introduces a buffer log based on parity logs to aggregate the data deltas generated by updates from multiple data blocks in the same position within the same stripe, thereby reducing the network traffic forwarded from the data blocks side to the parity log. However, CoRD's fixed-size buffer log does not account for read and write concurrency, causing the recycling process to become a bottleneck that limits update performance. In contrast, PL's extensive parity log space allows recycling to be indefinitely delayed without affecting update performance. The PL and FO mechanisms are two high-performance update methods; the random access advantage of SSDs makes FO's performance second only to PL. Similar to CoRD, PLR and PARIX exhibit suboptimal performance for two reasons: PLR integrates log recycle process into the update process, turning update appending operation into random operations due to numerous parity logs, while PARIX requires 2$\times$ network latency for update requests lacking temporal locality, which is particularly detrimental in a 25Gb/s cloud environment. 

Most importantly, these mechanisms all require a  time-consuming read-write process on the data block to compute data detlas. In contrast, TSUE introduces data logs at the data block side, moving the log recycle process to the backend for asynchronous real-time recycle, while the frontend handles the sequential appending of replica logs. Furthermore, TSUE employs a three-layer log architecture that leverages spatio-temporal locality to significantly reduce recycle overhead and designs a memory-based log pool structure, it supports concurrent log read and write operations.

% \begin{figure}[t]
%     \centering
%     \includegraphics[width=0.85\linewidth]{List-Update/F6-a.pdf}
%     \caption{Performance and memory overhead analysis.}
%     \label{fig:F7}
% \end{figure}

\subsection{Detailed Performance Analysis}
\label{performance analysis of TSUE}

\begin{figure}[!htb]
    \centering
    
    \begin{subfigure}{0.475\linewidth}
        \centering
        \includegraphics[width=\linewidth]{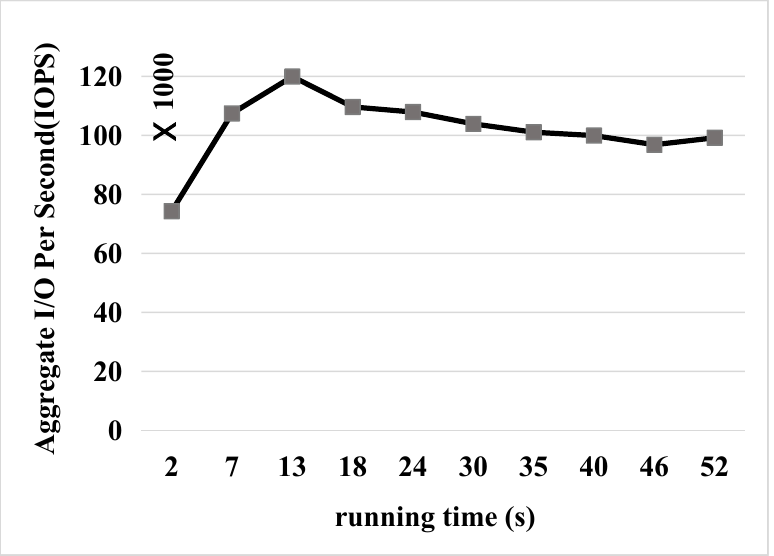}
        \caption{Recycle Overhead in Update}
        \label{fig:F7-01}
    \end{subfigure}
    \begin{subfigure}{0.49\linewidth}
        \centering
        \includegraphics[width=\linewidth]{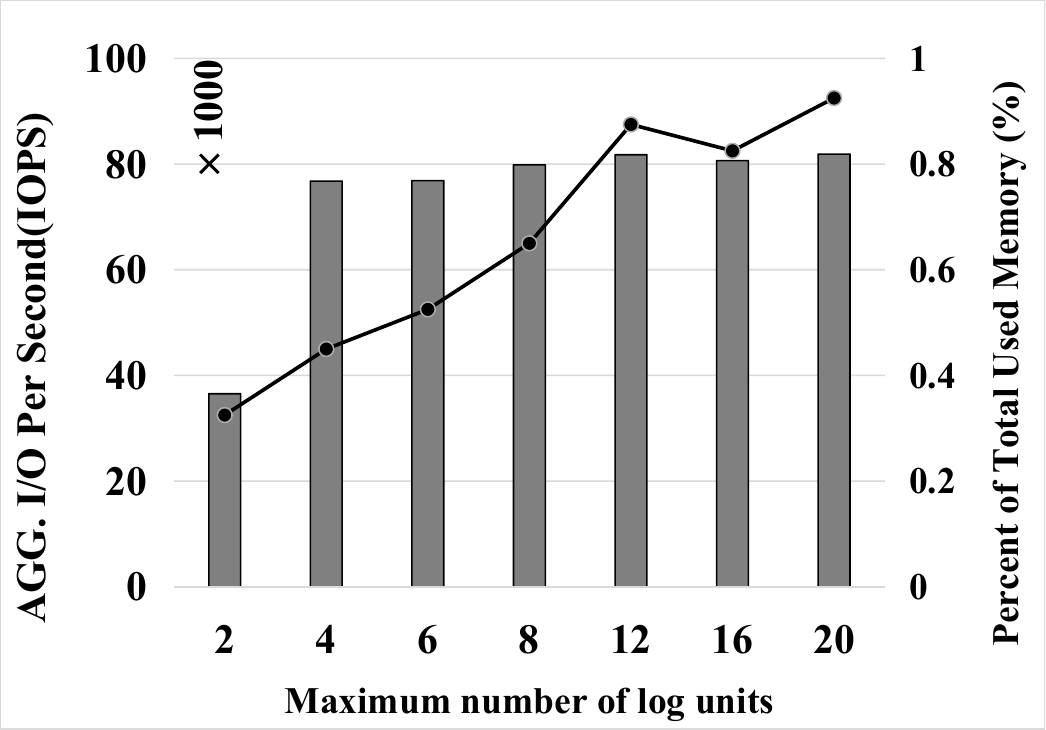}
        \caption{Memory Usage}
        \label{fig:F7-02}
    \end{subfigure}
    \caption{Performance and overhead analysis of TSUE.}
    \label{fig:F7}
%    \vspace{-3mm}
\end{figure}

\subsubsection{Log Recycle Overhead}
As shown in Fig. \ref{fig:F7-01}, we gather statistics on the update performance of TSUE over a one-minute period. When the maximum number of units is two, the update performance is minimal. Conversely, when the maximum number of units is set to four or more, the update performance becomes significantly higher and remains stable. The data indicates that the impact of the back-end log recycle process on update performance is negligible.

\subsubsection{Memory Usage}

TSUE dynamically adjusts the number of log units for each log pool—ranging from 2 to 20 in the data log, delta log, and parity log pools—based on access pressure. This mechanism enables adaptive expansion or release of log units. Each log unit is 16 MB, resulting in each log pool occupying between 32 MB and 320 MB of memory. TSUE configures four log pools for each of the data log, delta log, and parity log in each SSD. In addition, a copy of the data log in TSUE is exclusively stored in the SSD and is no longer retained in memory. Consequently, the total memory space configured by TSUE for each SSD ranges from 384 MB to 3840 MB. In Chameleon nodes, the memory footprint is between 0.15\% and 1.5\%. 

As shown in Fig. \ref{fig:F7-02},  we conducted an analysis of the update performance and maximum memory usage for varying log unit quotas in log pool structure. This indicates that, for optimal performance in memory-constrained environments, the maximum number of log units per log pool in TSUE is set to four.

Based on our testing statistics, we typically set the maximum number of log units to 4, enabling TSUE to reach peak performance with a maximum memory allocation of 1 GB for a single SSD.
 
\subsubsection{Performance Breakdown}

\begin{figure}[t]
    \centering
    \includegraphics[width=\linewidth]{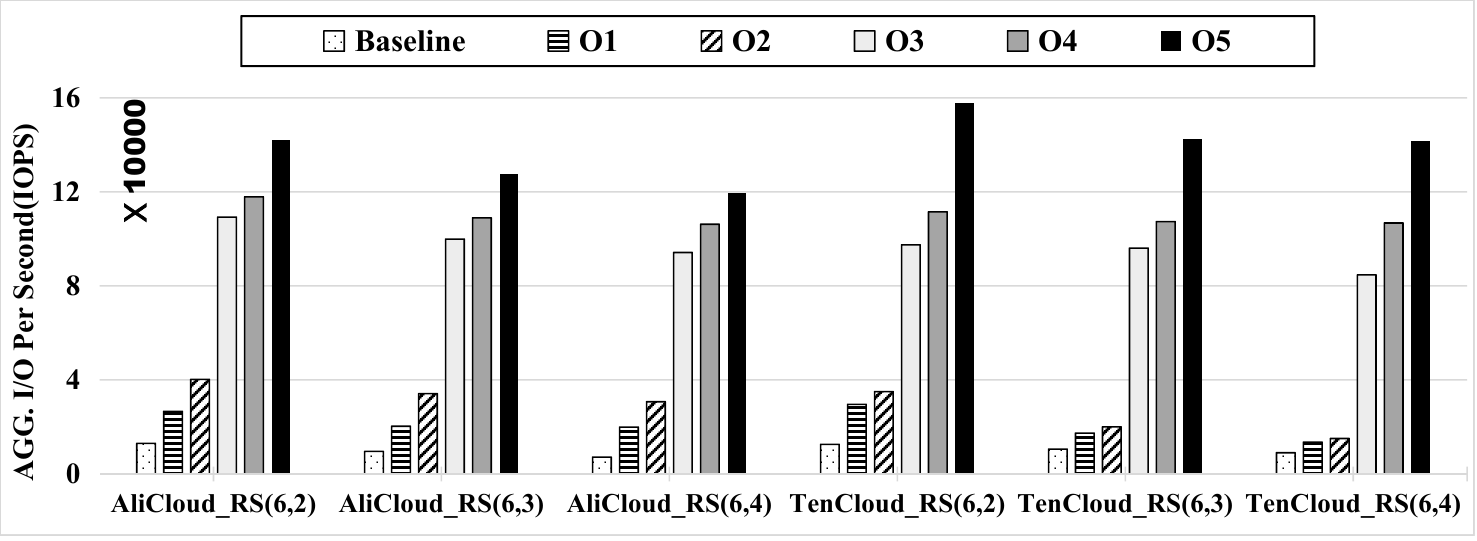}
    \caption{The breakdown analysis of update throughput.}
    \label{fig:F8}
\end{figure}

To better illustrate the contributions of this paper, we conducted a phased overlay test for each point mentioned, as shown in Fig. \ref{fig:F8}. Through the decomposition of TSUE, the baseline version of TSUE maintains two levels of logs: the data log and the parity log, both cached in memory to facilitate rapid appending and recycling. Subsequently, based on the baseline version, we gradually introduced the following enhancements:

\begin{itemize}
    \item \textbf{Baseline:} Utilizes DataLog and ParityLog to recycle data.
    \item \textbf{O1:} Utilizes spatio-temporal locality in data logs.
    \item \textbf{O2:} Utilizes spatio-temporal locality in parity logs.
    \item \textbf{O3:} Introduces log pool structure to manage log.
    \item \textbf{O4:} Configure 4 log pools for each SSD.
    \item \textbf{O5:} Introduces DeltaLog to reduce network load.
\end{itemize}

It is evident that Multi-LogPool (\textbf{O4}) contributes minimally to performance optimization. This suggests that in practical implementations involving a large number of SSDs or when memory resources are limited, assigning one log pool per SSD can significantly reduce memory usage without substantially compromising performance. As shown in the Fig. \ref{fig:F8}, the performance improvement achieved by \textbf{O1} is significantly higher than that of \textbf{O2}, highlighting the effectiveness of our optimization approach for the data block. The introduction of the log pool (\textbf{O3}) significantly enhances update performance by several times. This improvement is attributed to the specially designed structure of the log pool, which effectively leverages the locality characteristics of access patterns (\textbf{O1, O2}) to minimize the amount of data requiring recycling, thereby greatly improving update performance. In addition, the introduction of DeltaLog (\textbf{O5}) resulted in an approximate 30\% improvement in update performance. This enhancement is attributed to DeltaLog offloading the computational load of ParityLog recycling and employing the mechanism outlined in Equation (\ref{eq5}) to merge multiple data deltas for the same location across multiple data blocks within the same stripe, thereby reducing network load.

\subsubsection{I/O Workload}

In the update testing for SSD clusters, it is essential to consider not only performance metrics but also the I/O workload, write amplification caused by updates, and network traffic. As shown in Table \ref{table3}, TSUE exhibits the lowest number of read/write and overwrite operations among all methods. Furthermore, the network traffic generated during the update process in TSUE is only slightly higher than that of CoRD, which is specifically designed to minimize network traffic. Specifically, Specifically, read/write operations in TSUE is only 20\% of that in PL, 27\% of that in PARIX, and 94\% of that in CoRD. Similarly, TSUE’s overwrite operations are just 8\% of FO’s, 27\% of PARIX’s, and 37\% of CoRD’s. Due to their significantly lower read/write and overwrite counts, SSDs under the TSUE mechanism endure 2.5$\times$ - 13$\times$ longer than those using other update methods.

\begin{table}[h]
    \centering
    \renewcommand{\arraystretch}{1.5}
    \caption{Storage Workload and Network Traffic}
    \resizebox{\linewidth}{!}{
        \begin{tabular}{c c c c c c}
\toprule[0.5mm]
\multirow{2}{*}{\textbf{METHOD}} & \multicolumn{2}{c}{\textbf{READ/WRITE}} & \multicolumn{2}{c}{\textbf{OVERWRITE (Write Penalty)}} & \multirow{2}{*}{\textbf{\begin{tabular}[c]{@{}c@{}}NETWORK\\ TRAFFIC (GB)\end{tabular}}} \\ \cmidrule(lr){2-3} \cmidrule(lr){4-5} 
                                 & \textbf{Num.}    & \textbf{Volume (GB)} & \textbf{Num.}           & \textbf{Volume (GB)}         &                                                                                          \\ \midrule[0.3mm]
FO                      & 9,739,320        & 894                  & 4,869,660                & 447                          & 447                                                                                      \\
PL                      & 13,726,649       & 1,252                & 4,869,660                & 447                          & 447                                                                                      \\
PLR                     & 8,860,540        & 1,240                & 7,493,637                & 440                          & 447                                                                                      \\
PARIX                   & 6,439,527        & 471                  & 1,456,968                & 96                           & 454                                                                                      \\
CoRD                    & 2,913,936        & 365                  & 1,094,691                & 271                          & 276                                                                                      \\
\hline
\textbf{TSUE}                    & \textbf{2,750,796} & \textbf{645}         & \textbf{401,466}         & \textbf{223}                 & \textbf{290}                                                                             \\ \bottomrule[0.5mm]
\end{tabular}
    }
    \begin{minipage}{0.94\linewidth}
        \vspace{1mm}
        \footnotesize
        \raggedright
        $\bullet$\ \textbf{NOTE:} Replaying Ten-Cloud Trace in RS(6,4). The data log and delta log in TSUE use the dual-copy mechanism, and all the logs are persisted to SSDs.
    \end{minipage}
    \label{table3}
\end{table}

It can also be observed that the total read/write volume of TSUE is higher than that of PARIX and CoRD, with TSUE's total overwrite volume being higher than PARIX's, and its network traffic slightly higher than CoRD's. To enhance log recycle performance, TSUE implements a three-layer log structure for efficient log management. In contrast, other update methods write only a party log during the update process. In addtion, the data in parity log of PARIX is not recycled in real-time as in TSUE, the original data in the storage device for repeated update locations will not be read twice.

\subsubsection{Time of Updated Data Resided in Memory on Average:}

As shown in Table \ref{table2}, the average residence time of a request in memory is around  10 seconds, which means the average time from data update to its merging into the corresponding data block and parity blocks is about 10 seconds. To ensure sufficient reliability within this short duration, we utilize double-copy data logs and delta logs.

\begin{table}[h]
    \centering
    \renewcommand{\arraystretch}{1.4}
    \caption{Time (in us) of Data Resided in Memory}
    \setlength{\tabcolsep}{6pt}
    \resizebox{\linewidth}{!}{
        \begin{tabular}{cc c c c c}
\toprule[0.5mm]
\multicolumn{2}{c}{\textbf{TRACE}} & \textbf{DATA\_LOG} & \textbf{DELTA\_LOG} & \textbf{PARITY\_LOG} & \textbf{TOTAL TIME} \\ 
\midrule[0.3mm]
\multirow{3}{*}{\textbf{Ali-Cloud}} & \textbf{APPEND} & 107 & 75 & 136 & \multirow{3}{*}{\textbf{9151364}} \\ 
                                      & \textbf{BUFFER} & 1561410 & 6846260 & 742066 &  \\ 
                                      & \textbf{RECYCLE} & 350 & 526 & 434 &  \\ 
\cline{1-6}
\multirow{3}{*}{\textbf{Ten-Cloud}} & \textbf{APPEND} & 96 & 185 & 218 & \multirow{3}{*}{\textbf{10979659}} \\ 
                                      & \textbf{BUFFER} & 4388070 & 59434880 & 64447 &  \\ 
                                      & \textbf{RECYCLE} & 323 & 237 & 2363 &  \\ 
\bottomrule[0.5mm]
\end{tabular}

    }
    \label{table2}
    \begin{minipage}{0.94\linewidth}
        \vspace{1mm}
        \footnotesize
        \raggedright
        $\bullet$\ \textbf{NOTE:} The Execution is performed under RS(12,4).
    \end{minipage}
\end{table}

It can be observed that the time for log append and recycle is on the order of microseconds or milliseconds. The maximum interval from when the log unit receives data until it is reclaimed is 7 seconds, while the minimum interval is 64 milliseconds. Reducing the log size from 16MB to 8MB can halve this interval, thereby allowing the log size to be adjusted based on the residence time. Additionally, increasing network performance from 25Gb/s to 100Gb/s or 200Gb/s can further reduce this time.

\begin{comment}
    \subsection{TSUE with Different Storage Devices}
    \subsubsection{TSUE with Low-speed Devices}
\end{comment}
\subsection{TSUE with HDDs}
\begin{comment}
    As can be observed from Fig. \ref{fig:F6}, \ref{fig:F7} and \ref{fig:F8}, with the increase in the number of clients, the performance of TSUE initially rises and subsequently declines. There are several concentrated reasons for this phenomenon, encompassing data placement issues, hot spot problems, as well as network and disk matters.
\end{comment}

\label{low-speed-test}
We established a 16-node test cluster, each node equipped with 32 GB memory, three 2 TB HDDs, and interconnected via 40 Gb/s InfiniBand switches.  A log pool structure is configured for each HDD device. Given the relatively poor performance of HDDs, only data logs and parity logs are enabled, while delta logs are omitted to optimize network traffic. 

\begin{figure}[!htb]
    \centering
    \begin{subfigure}{0.49\linewidth}
        \centering
        \includegraphics[width=\linewidth]{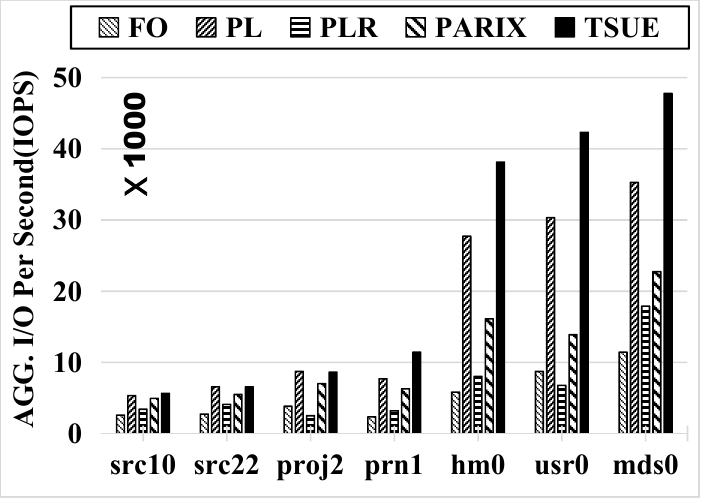}
        \caption{Update Throughput}
        \label{fig:F11-01}
    \end{subfigure}
    \begin{subfigure}{0.49\linewidth}
        \centering
        \includegraphics[width=\linewidth]{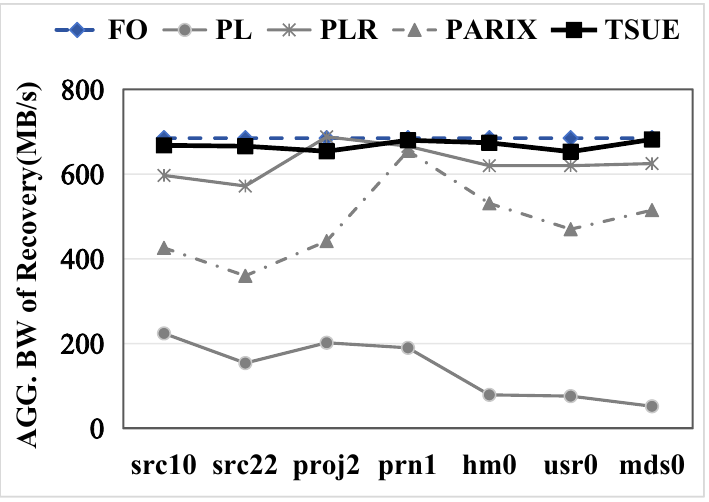}
        \caption{Recovery Bandwidth}
        \label{fig:F11-02}
    \end{subfigure}
    \caption{Performance evaluation of TSUE with HDDs.}
    \label{fig:F11}
%    \vspace{-3mm}
\end{figure}

The MSR-Cambridge trace is selected for comparison. We conducted extensive testing in HDD environments. Due to space constraints, we briefly describe RS (6,4) as an example. As shown in Fig. \ref{fig:F11-01}, the aggregation IOPS of TSUE with RS(6,4) is up to 3.6$\times$ that of PARIX, 9.1$\times$ that of PLR, 4$\times$ that of PL and 16.2$\times$ that of FO. The performance of TSUE is best in all SOTA methods for HDDs cluster. In addition, we conducted data recovery tests in the identical environment, we initiated the data recovery test immediately after terminating all client update requests. The update test had been running for 3 minutes prior to this action, as scripted, to verify the impact of the logs generated during the update process on recovery performance. As shown in Fig. \ref{fig:F11-02}, the recovery performance of TUSE is most similar to that of the FO mode without logs. Additionally, the impact of logs on recovery in TSUE can be considered negligible because the logs in TSUE are recycled in real-time.

\begin{comment}
\subsubsection{TSUE with High-speed Devices vs the Low-speed Devices:} 
\end{comment}

The comparison between Section  \ref{high-speed-test} and Section \ref{low-speed-test} demonstrates that an SSD cluster equipped with a 25 Gb/s network achieves significantly higher performance than an HDD cluster equipped with a 40 Gb/s network. In the update mechanism, the workload on storage typically imposes approximately twice the network workload. Therefore, it is crucial to prioritize improving storage device performance, which yields the most substantial performance gains, before optimizing network performance.

\section{Related Work}
The erasure code update mechanism involves a lengthy update path and introduces additional random access operations, network load, and computational overhead, thereby posing significant challenges for optimizing update performance. In recent years, the incremental update mechanism has emerged as the primary method to address these challenges. Below are some key works in this area.

First, to address the additional disk access and network load introduced by reconstruct updates, FO \cite{ref58} adopt in-place update method for updating data block and parity blocks, however, all operations during the update process are small-grained random accesses, which significantly degraded the performance of early HDD-based storage systems. To mitigate the random access issue, FL \cite{ref1,ref11} employs log to record all data and parity deltas, however, it has a significant negative impact on data reading performance.

Second, PL \cite{ref6} adopts an in-place update method for data blocks and utilizes parity logs to store parity deltas. However, the recycling of parity logs is only triggered when disk usage reaches a certain threshold, potentially compromising data consistency. To address this issue, PLR \cite{ref2} reserves space adjacent to the parity blocks for storing corresponding parity deltas, thereby avoiding the random-accesses associated with log recycling. Nonetheless, the presence of a large number of log files can make log appending operations resemble random writes, and the performance of log appending is limited by the log recycling process.

Thirdly, PARIX \cite{ref3, ref59} proposes a speculative update method that records both updated and original data in the parity log to avoid the time-consuming write-after-read process required to calculate parity deltas. For access patterns exhibiting temporal locality, this approach ensures that the original data is read only once, thereby reducing data reading costs. However, this method focuses primarily on temporal locality and may still benefit from optimization due to its local update mode of data blocks.

Fourth, CoRD \cite{ref55} combines data deltas from multiple data blocks within the same stripe to reduce network traffic from data block side to the parity log. Additionally, it introduces a technique for flipping even data blocks to exploit spatial locality further, which is not suitable for environments with large data blocks. Moreover, CoRD was not designed to handle update concurrency effectively.

In addition,  several works have optimized the recovery and update processes of erasure codes from different perspectives, thereby complementing our approach. LASC \cite{ref57}, a locality-aware speculative cache scheme for partial updates, identifies the update locality of a data block from an update request stream over a given period and speculatively caches the old data from the disk. By caching entire old data blocks with high update locality, it reduces disk seeks. This approach effectively mitigates the seek overhead associated with HDDs; however it introduces additional read overhead in SSDs. LogECMem \cite{ref28}  maintains data blocks and the first XOR parity chunk of stripe in DRAM and stores non-XOR parity chunks in disk nodes to reduce memory overhead. It performs in-place updates for data and XOR parity chunk in RAM, and utilizes parity logging for non-XOR parity chunks on disk. However, it is specific to Memcached and not applicable to traditional file systems where data blocks and parity blocks are stored on disk.

In summary, while existing work has optimized the updating process from various aspects, there is still significant room for improving update performance. TSUE leverages spatiotemporal locality to substantially reduce the actual number of requests through its multi-level log structure, offering a novel solution for update performance optimization.

\section{Conclusion and Future Work}

This study presents TSUE, a two-stage data update method that leverages spatial-temporal locality to significantly reduce random-access, network, and computational overheads while enabling real-time recycling of data logs, delta logs and parity logs, and minimize the negative impact of log reclamation on data recovery in data loss scenarios. In comparative tests against methods like FO, PL, PLR, PARIX and CoRD, TSUE consistently achieved the highest aggregation IOPS and lowest latency.

In the future, we plan to deeply optimize the erasure code update process, particularly for recycle efficiency of log.  Moreover, we aim to further enhance the efficiency of erasure code updates by integrating high-bandwidth networks such as InfiniBand, high-performance storage devices like NVM, and accelerators like GPUs. 

Furthermore, we explored the integration of compression mechanisms into the update process to alleviate network traffic congestion. As shown in Table \ref{table2}, The log content remains in each layer of the logging system for approximately 1 to 5 seconds. This duration is adequate to facilitate the compression and decompression processes, thereby further reducing network traffic and enhancing overall throughput.

\section*{Acknowledgment}
The work is supported by the National Natural Science Foundation of China (Grant Nos. 62032023, T2125013, and 61502454), the National Key Research and Development Program of China (Grant No. 2025YFB30037002), and the Innovation Funding of ICT, CAS (Grant No. E461050).

\newpage

\bibliographystyle{ACM-Reference-Format}
\bibliography{tsue.bib}

\end{document}